\documentclass[12pt]{article}

\usepackage{mathpartir}
\newcommand{\annot}[1]{\left\{\begin{aligned}#1\end{aligned}\right\}}
\newcommand{\lambdaFn}[2]{(\lambda #1.\; #2)}
\newcommand{\goesto}{\rightarrowtail}

\newcommand{\hoareTriple}[3]{\annot{#1} \; #2 \; \annot{#3}}
\newcommand\sepimp{\mathrel{-\mkern-6mu*}}

\usepackage{dashbox}
\newcommand\dboxed[1]{\dbox{\ensuremath{#1}}}
\def\checkmark{\tikz\fill[scale=0.4](0,.35) -- (.25,0) -- (1,.7) -- (.25,.15) -- cycle;} 
\usepackage[T1]{fontenc}
\usepackage{textcomp}
\usepackage{lmodern}

\usepackage[nottoc]{tocbibind}

\RequirePackage{natbib}
\renewcommand\cite{\citep}	
\newcommand*\set[1]{\left\{#1\right\}}
\newcommand{\code}[1]{\textnormal{\texttt{#1}}}
\usepackage[export]{adjustbox}
\newcommand{\refCommand}{\text{\texttt{\textbf{ref}}}}
\newcommand{\forkCommand}[1]{\text{\texttt{\textbf{fork}}}\;\ensuremath{\set{#1}}}
\newcommand{\letCommand}[3]{\text{\texttt{\textbf{let}}}\;\ensuremath{#1}\code{ = }\ensuremath{#2}\;\text{\texttt{\textbf{in}}}\;\ensuremath{#3}}
\usepackage{eucal}
\newcommand{\mval}{\mathcal{V}}
\newcommand{\subsubsubsection}[1]{\paragraph{#1}\mbox{}\\}
\usepackage{float}
\usepackage{bussproofs}
\usepackage{xcolor}
\usepackage[colorlinks = true,
            linkcolor = blue,
            urlcolor  = blue,
            citecolor = blue,
            anchorcolor = blue]{hyperref}

\newcommand{\MYhref}[3][blue]{\href{#2}{\color{#1}{#3}}}%

\newcommand*{\savelabel}[2]{%
	\edef\@currentlabel{#2}
	\phantomsection
	\label{#1}
}
\newcommand*{\textlabel}[2]{{#2}\savelabel{#1}{#2}}

\def\optionaldischarge#1{%
	\if\relax\detokenize{#1}\relax\else\ensuremath{^{#1}}\fi}

\newcommand*{\inferhref}[4]{%
	\inferrule*[lab=\textlabel{#2}{#1}]{#3}{#4}%
}

\newcommand*{\biimp}[2]{%
	\hbox{%
		\ooalign{%
			$\genfrac{}{}{1.6pt}1{#1}{#2}$\cr%
			$\color{white}\genfrac{}{}{0.8pt}1{\phantom{#1}}{\phantom{#2}}$%
		}%
	}%
}
\newcommand{\BIIMP}{\mprset{myfraction=\biimp}}

\newcommand*{\inferH}[3]{\inferhref{#1}{#1}{#2}{#3}}

\newcommand*{\inferHB}[3]{{\BIIMP\inferH{#1}{#2}{#3}}}

\usepackage{proof}
\usepackage{hyperref}

\usepackage[a4paper]{geometry}

\newcommand{\val}{v}
\usepackage[utf8]{inputenc} 
\usepackage[T1]{fontenc} 

\usepackage{times}

\usepackage{microtype}  
\usepackage{inconsolata} 
\usepackage[english]{babel} 

\usepackage{amsmath} 
\usepackage{amsthm}
\usepackage{amssymb}

\usepackage[labelsep=period]{caption}

\usepackage{array}
\usepackage{tabu}   
\usepackage{xspace} 

\usepackage{graphicx}
\graphicspath{{figures}}

\usepackage[all]{hypcap}

\usepackage{color}
\usepackage{xcolor} 
\definecolor{mauve}{rgb}{0.58,0,0.82}

\let\chapter\section
\usepackage[ruled, vlined, linesnumbered]{algorithm2e}

\SetKw{True}{true}
\SetKw{False}{false}
\SetKwData{typeInt}{Int}
\SetKwData{typeRat}{Rat}
\SetKwData{Defined}{Defined}
\SetKwFunction{parseStatement}{parseStatement}

\usepackage{todonotes}

\usepackage{verbatim}

\usepackage{listings}
\usepackage{proof}
\usepackage{semantic} 
\def\predicatebegin #1\predicateend{$\Gamma \vdash #1$}

\usepackage{mathtools}

\usepackage{stackengine}
\usepackage{scalerel}
\usepackage{fp}
\def\clipeq{\kern -.65pt\mathrm{=}\kern -1pt}
\def\Lla{\rule[-.1ex]{.3pt}{1.3ex}}
\def\Lra{\kern -3.5pt\Longrightarrow}
\newcount\argwidth
\newcount\clipeqwidth
\savestack\tempstack{\stackon{$\clipeq$}{}}%
\clipeqwidth=\wd\tempstackcontent\relax

\newcommand{\spac}{\hskip 0.2em plus 0.1em} 
\def\Ret #1.{#1.\spac}%

\newcommand\xMapsto[2]{%
  \savestack\tempstack{\stackon{$\scriptstyle#1$}{$\scriptstyle#2$}}%
  \argwidth=\wd\tempstackcontent\relax%
  \FPdiv\scalefactor{\the\argwidth}{\the\clipeqwidth}%
  \FPsub\scalefactor{\scalefactor}{1.5}
  \FPmax\scalefactor{\scalefactor}{.05}%
  \mathrel{%
  \stackunder[2pt]{\stackon[3pt]{$\Lla\hstretch{\scalefactor}{\clipeq}\Lra$}%
     {$\scriptstyle#1~$}}{$\scriptstyle#2~$}%
  }%
}

\begin{document}

\thispagestyle{empty}
\begin{center}
\thispagestyle{empty}
\large
INDIANA UNIVERSITY BLOOMINGTON\\

\vspace{25mm}

\Large Elizabeth Dietrich

\vspace{4mm}

\huge A beginner's guide to Iris, Coq and separation logic

\vspace{20mm}

\Large
\iflanguage{english}{
Bachelor's Thesis
}

\end{center}

\vspace{2mm}

\begin{flushright}
 {
 \setlength{\extrarowheight}{5pt}
 \begin{tabular}{r l} 
  \sffamily \iflanguage{english}{Supervisor(s)}{}: &                    \sffamily Dennis Shasha \\
 \end{tabular}
 }
\end{flushright}


\vfill
\centerline{\large\the\year}


\newpage

\newcommand\EngInfo{{%
\selectlanguage{english}
\noindent\textbf{\large  A beginner's guide to Iris, Coq and separation logic}

\vspace*{3ex}

\noindent\textbf{Abstract:}
Creating safe concurrent algorithms is challenging and error-prone. For this reason, a formal verification framework is necessary especially when those concurrent algorithms are used in safety-critical systems. The goal of this guide is to provide resources for beginners to get started in their journey of formal verification using the powerful tool Iris.  The difference between this guide and many others is that it provides (i) an in-depth explanation of examples and tactics, (ii) an explicit discussion of separation logic, and (iii) a thorough coverage of Iris and Coq. References to other guides and to papers are included throughout to provide readers with resources through which to continue their learning.

\noindent

\vspace*{1ex}

\vspace*{1ex}

\vspace*{1ex}
}}

\iflanguage{english}{\EngInfo}{\EstInfo}

\newpage
{
  \hypersetup{linkcolor=black}
  \tableofcontents
}

\newpage
\section{Introduction}
Iris is a higher-order separation logic framework for reasoning about concurrent programs.
This guide explains how to use Iris in formal verification. We refer to formal verification as a way to use mathematical techniques to prove, in a rigorous and machine-checkable manner, the absence of bugs and the conformity of a system to its intended specification. 



Unfortunately, Iris is  difficult to learn. This guide attempts to teach you how to approach an Iris proof on your own. We assume no prior knowledge of Hoare logic or Coq and provide a section on separation logic to help build your intuition. This guide should serve as a starting point to your journey in learning Iris and provides links to resources throughout the sections for you to delve deeper into formal verification. 

\subsection{Case Studies}\label{intro-case} 
In Section \ref{beginning-example} we implement a counter that increments by one in HeapLang, a simple default language for Iris. HeapLang is a functional language with mutable references that allows us to write directly from Coq \cite{mit}. We will delve more into the specifics of HeapLang in Section \ref{heaplang}. We break the proof down into its different components that will allow us to gain intuition about Iris and separation logic. We show two different implementations of this counter. The first implementation is a simple program that will be explained in-depth in Section \ref{sep-formulas} and the second implementation will be a more advanced parallel increment example. This parallel increment example is fully explained in the \MYhref{https://gitlab.mpi-sws.org/iris/examples/-/blob/master/theories/lecture_notes/coq_intro_example_1.v}{example files} of the Iris Project lecture notes. Therefore, we will not re-explain the example in its entirety here, but we will highlight its invariant. 

The case study example will evolve into  a more advanced bank example which allows the transfer of money between two bank accounts.  The balance of these accounts will be represented as mathematical integers and their sum will be 0. Before we allow transfer from one account to the other we will need to ensure that the accounts have a sum of 0. 
The bank example will be fully explained in Section \ref{case-work}. 

A more general bank example is taken from the guide: \MYhref{https://plv.csail.mit.edu/blog/iris-intro.html}{"A Brief Introduction to Iris"}. 
For readers looking to further their expertise after working through this guide, we suggest that as a starting point. 

These examples are meant to be easy enough for you to try to re-implement on your own after working through this guide. For more examples, many of which are advanced, refer to the \MYhref{https://gitlab.mpi-sws.org/iris/examples/-/tree/master/theories/lecture_notes}{Iris Project lecture notes}. 

\subsection{Summary and Outline}
This guide explains the verification tools Iris and Coq, along with the basics of separation logic necessary to understand how to use these tools. 
\begin{itemize}
    \item Section \ref{installing} covers how to install all of the software we use throughout this guide including opam, Coq, and Iris. 
    \item Section \ref{coq} introduces the Coq proof assistant. We present a simple proof to introduce the basics of proof mode, the type of reasoning Coq utilizes, and some of the main tactics available. 
    \item In order to fully understand what is happening in Iris,  we introduce the basics of separation logic in Section \ref{sepLogic}.  
    \item Before we dive into the uses of Iris, we take a brief look at the base logic of Iris in Section \ref{base-logic}. Here we further explain our simple increment example using proof rules and Hoare logic.
    \item Section \ref{iris-discussion} introduces the notation and some basic tactics of Iris. It presents how each tactic works and provides some examples on how a tactic might affect the state of your proof goal.
    \item Section \ref{set-up} covers what is required to build your own proof in Iris from the ground-up.
\end{itemize}
This guide aims to be accessible to readers who have no prior background in formal verification. We hope that this tutorial on the Iris and Coq verification tools will allow you to start your journey into formal verification using these tools for proof mechanization. Further resources are provided throughout this guide to provide you with some possible next steps in learning more about Iris and Coq. 

\newpage 
\section{Installing the necessary software} \label{installing}
\subsection{Installing Opam}\label{opam}
Before you can install Iris and Coq, you must ensure you have opam (version 2.0.0 or newer) installed on your device. Opam is the package manager for OCaml, the programming language in which Coq is implemented, that supports multiple simulataneous compiler installations and flexible package constraints. In order to install opam using Homebrew use command: \framebox{brew install opam}. Once opam is installed, it must be intialized before you can use it. We do this by the following commands in order: \framebox{opam init}, \framebox{eval \$(opam env)}. \framebox{opam init} will prompt you to allow opam to set up intialization scripts which is generally fine to accept. If you do not accept, every time a new shell is opened, you will have to type the \framebox{eval \$(opam env)} command to update environment variables. For further help with installing opam, including different installation avenues, refer to \MYhref{https://opam.ocaml.org/doc/Install.html}{the opam documentation}. 

\subsection{Installing Coq}
While we briefly mention how to install opam above, the Coq platform provides interactive scripts that allow installing Coq and its packages through opam without having to learn anything about opam. You can use the command, \framebox{opam install coq} to install Coq on your device. Depending on your operating system, installing Coq using opam may require you to first install system packages. You can check the \MYhref{https://coq.inria.fr/opam-using.html}{guide to installing Coq} to see your personal device needs, here you will also find how to install the CoqIDE and other Coq packages. 

\subsection{Installing Iris}\label{install-iris}
In this guide we refer to Iris implemented and verified in the Coq proof assistant. Therefore, in terms of installation we will present the Coq development of the Iris Project, a general proof mode for carrying out separation logic proofs in Coq. A more in-depth guide can be found \MYhref{https://gitlab.mpi-sws.org/iris/iris}{here}. To use Iris in your own proofs, you should install Iris via opam (2.0.0 or newer). In order to obtain the latest stable release, you have to add the Coq opam repository as follows: 
\begin{center}
\framebox{
opam repo add coq-released https://coq.inria.fr/opam/released}
\end{center}
Now you can install Iris:
\begin{itemize}
    \item \framebox{opam install coq-iris} will install the libraries making up the Iris logic, but leave it up to you to instantiate the program\_logic.language interface to define a programming language for Iris to reason about. 
    \item \framebox{opam install coq-iris-heap-lang} will additionally install HeapLang, the default language used by various Iris projects. 
\end{itemize}
This guide assumes both commands are called by the user. To fetch updates later, run \framebox{opam update} and \framebox{opam upgrade}. 
\newline \newline 
Once you have Iris installed, you can set up your editor to properly input and output the unicode characters used throughout Iris. While this guide will not provide instructions, please refer to this in-depth \MYhref{https://gitlab.mpi-sws.org/iris/iris/-/blob/master/docs/editor.md}{guide} for further instructions. However, we do include an Appendix to serve as a guide for inputting Unicode symbols that applies to Visual Studio Code with the 'Generic Input Method' extension installed and configured as described in the guide previously mentioned.

\newpage
\section{What is Coq}\label{coq}
Coq is a formal proof management system. It provides a rich dependently-typed framework to formalize mathematics and programming language metatheory. In order to make proofs feasible to construct, Coq supports a range of built-in \emph{tactics} which are engineered primarily to support \emph{backward reasoning}. When programming in Coq, you can use these tactics to manipulate the proof state interactively, applying axioms or lemmas to break the goal into subgoals until all subgoals have been solved. This allows Coq to infer many details about how to instantiate lemmas by inspection of the current proof state, and as a result the Coq user can omit these details from their interactive proof scripts. 

\subsection{Basics of proof writing}
Coq is an interactive proof assistant, which means that proofs can be constructed interactively through a dialog between the user and the assistant. The building blocks for this dialog are tactics that the user can use to represent steps in the proof of a theorem. 

\subsubsection{Proof Mode}
Proof mode is used to prove theorems. It is the core mechanism of the dialog between the user and the proof assistant. Coq enters proof mode when you begin a proof, such as with the \textcolor{blue}{\textbf{Theorem}} or \textcolor{blue}{\textbf{Lemma}} command, and it exits proof mode when you complete a proof, such as with the \textcolor{blue}{\textbf{Qed}} command. Tactics, which are available only in proof mode, incrementally transform incomplete proofs to eventually generate a complete proof and will be discussed more below in Section \ref{tactics-coq}. When you run Coq interactively through the CoqIDE or coqtop, which will be shown later, Coq shows the current proof state (the incomplete proof) as you enter tactics. 

\subsubsubsection{Proof State}\label{proof-state}
The proof state consists of one or more unproven goals. Each goal has a conclusion, which is the statement that needs to be proven, and a local context, which contains named hypotheses (which are propositions), variables, and local definitions that can be used in proving the conclusion. The proof may also use constants from the global environment such as definitions and already proven theorems. The term goal may refer to either an entire goal or to the conclusion of a goal, depending on the context.

The conclusion appears below a line and the local context appears above the line. The local context of a goal contains items specific to the goal as well as section-local variables and hypotheses defined in the current section. The latter are included in the initial proof state. Items in the local context are ordered; an item can  refer only to items that appear before it. The global environment has definitions and proven theorems that are global in scope. 
\begin{figure}[H]
\centering
\begin{minipage}[t]{.65\linewidth}
\begin{lstlisting}[mathescape=true]
Lemma example: $\forall$ b: bool, b=true $\lor$ b=false.
Proof.
intros. destruct b. left. done. right. done.
Qed.
\end{lstlisting}
\end{minipage}
\caption{Proof that every Boolean is either true or false.}
 \label{fig:example1proof}
\end{figure}

\noindent Let us look at the example lemma in Figure \ref{fig:example1proof} that will prove every Boolean is either true or false. When you begin proving a lemma, the proof state shows the statement of the theorem below the line and often nothing in the local context, as seen in Figure \ref{fig:bool}. 
\begin{figure}[H]
  \includegraphics[width=150mm]{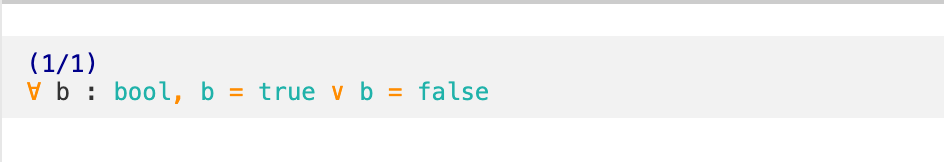}
  \centering
  \caption{After defining the Boolean lemma, we use tactic $\textcolor{blue}{\textbf{Proof}}$ to start our proof. This gives us the proof state of the theorem we must now prove.}
  \label{fig:bool}
\end{figure}

\noindent After applying the $\textcolor{blue}{\textbf{intros}}$ tactics, we see the hypothesis: $\scriptsize{b = true \lor b = false}$. This also introduces the term "b" into our local context. The name of the variable (b) appears before the colon, followed by the type (bool) that it represents which can be seen in Figure \ref{fig:bool-step1}. 
\begin{figure}[H]
  \includegraphics[width = 150mm]{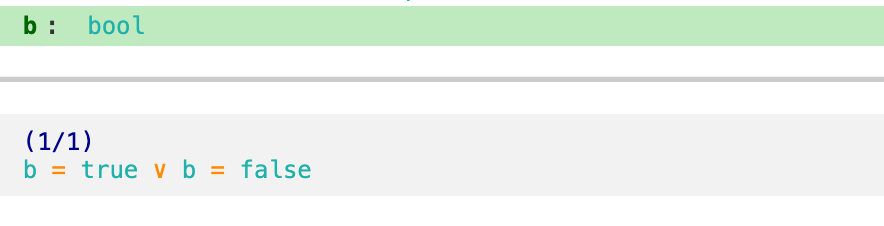}
  \centering
  \caption{After using the tactic $\textcolor{blue}{\textbf{intros}}$ we need to prove $b = true \lor b = false$ and the term "b" was added to our local context.}
  \label{fig:bool-step1}
\end{figure}

\noindent "Variables" may refer specifically to local context items and "hypotheses" refers to items that are propositions. However, these terms are often used interchangeably. We use the knowledge that b is a bool by calling the tactic $\textcolor{blue}{\textbf{destruct}}$ in Figure \ref{fig:bool-intro} which will split the proof according to the two cases. 

\begin{figure}[H]
  \includegraphics[width = 150mm]{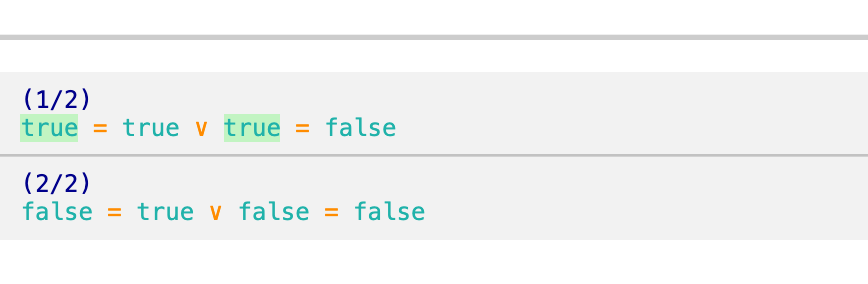}
  \centering
  \caption{The tactic $\textcolor{blue}{\textbf{destruct b}}$ splits the proof according to the two cases and takes b from our local context to show all of the possibilities of its value. This is an example of a state with two subgoals.}
  \label{fig:bool-intro}
\end{figure}

\noindent In each case, it is easy to determine that both $\scriptsize{true = true}$ and $\scriptsize{false = false}$ is trivial. This means we can finish the proof off by trivially showing the left or right hand side of each case is true. In order to do so, we can use the tactics  $\textcolor{blue}{\textbf{left}}$ and $\textcolor{blue}{\textbf{right}}$ for each respective side. These tactics will take the correct side of our logical disjunction. For example, $\textcolor{blue}{\textbf{left}}$ will result in the statement $\scriptsize{true = true}$ of our first subgoal. After we use these tactics, we can call the tactic  $\textcolor{blue}{\textbf{done}}$ on both cases since this statement is trivially correct. We then finish our proof and exit proof mode through the $\textcolor{blue}{\textbf{Qed}}$ tactic. Coq's kernel verifies the correctness of proof terms when it exits proof mode by checking that the proof term is well-typed and that its type is the same as the theorem statement. The full proof involving these tactics can be seen in Figure \ref{fig:example1proof}. When we write "proof" we refer to a proof script that consists of the tactics that are applied to prove a theorem.

\subsubsubsection{Tactics}\label{tactics-coq}
Tactics are available only in proof mode. They specify how to transform the proof state of an incomplete proof to eventually generate a complete proof. Coq and its tactics use backward reasoning. In backward reasoning, the proof begins with the theorem statement as the goal which is then gradually transformed until every subgoal generated along the way has been proven. For example, take the proof of $A \land B$. This proof would begin the formula $A \land B$ as the goal. This can then be transformed into two subgoals $A$ and $B$, followed by the proofs of $A$ and $B$. 

A tactic may fully prove a goal, in which case the goal is removed from the proof state. However, more commonly, a tactic replaces a goal with one or more subgoals. Most tactics require specific elements or preconditions to reduce a goal and will display an error message if the tactic cannot be applied to the goal. Tactics are applied to the current goal by default. 

While a few tactics were discussed in the example above, we will not delve deeper into the specifics of Coq tactics in this guide, as we focus on using Iris tactics as found in Section \ref{IPM}. An in-depth guide to Coq tactics can be found in \MYhref{https://coq.inria.fr/refman/index.html}{the reference manual of Coq}.

\newpage
\section{Separation Logic} \label{sepLogic}
In this user guide, we use \emph{separation logic} to specify and verify concurrent data structures. We based this section largely off of the text of \cite{book}). Separation logic (SL), an extension of Hoare logic, is used for reasoning about programs that access and mutate data held in computer memory. SL is based on the separating conjunction \(P * Q\), which asserts that \emph{P} and \emph{Q} hold for separate portions of memory and allows one to modularly describe states of a program through proof rules that rely on separation \cite{spl}. Each sentence in this language is called a \emph{formula}, and describes a set of states. We say a state \emph{satisfies} a formula when the state is described by the formula. The set of proof rules that SL gives us can be used to prove that states of interest (such as the set of resulting states after a program executes) satisfy a particular formula. SL can be used in many ways including automatic program-proof tools or abstract interpreters.

Here, we will look at SL's relationship to \emph{Iris}, a framework for higher-order concurrent separation logic, implemented in the Coq proof assistant. While this guide will touch only on the basics of SL, the core Iris logic is very abstract. We refer readers to the paper \MYhref{https://www.cambridge.org/core/services/aop-cambridge-core/content/view/26301B518CE2C52796BFA12B8BAB5B5F/S0956796818000151a.pdf/iris_from_the_ground_up_a_modular_foundation_for_higherorder_concurrent_separation_logic.pdf}{Iris from the ground up} for a more in-depth look into this logic. 

\subsection{Basics}
In this section we will introduce the basic constructs of SL before we consider the formal definitions and formulas that we will need to understand Iris.

\subsubsection{"Points-To" Connective}\label{sep-con}
Separation logic is a \emph{resource} logic; not only do propositions denote facts about the state of the program, but also \emph{ownership} of resources. In this section, we fix the notion of a resource to be a heap, a piece of global memory represented as a finite partial mapping from memory locations to the values stored there. The \emph{points-to} connective, $p \mapsto q$, asserts that the program state contains a heap cell at address $p$ that contains value $q$. If we are verifying an expression \emph{e} that has $p \mapsto q$ in its precondition, then we can assume not only that the location \emph{p} currently points to \emph{q}, but also that the \emph{right} to update \emph{p} is \emph{owned} by \emph{e}. Ownership, in this case, means that when verifying \emph{e} we would not need to consider the possibility that another piece of code in the program might interfere with \emph{e} by updating \emph{p} during \emph{e}'s execution \cite{iris-ground-up}. We will further elaborate on this on the next page. 

\subsubsection{Separating Conjunction}
The separating conjunction, $*$ (also known as \emph{star}), allows us to state properties in disjoint parts of the memory. Unlike the standard conjunction $\land$, the separating conjunction conjoins two formulas that describe \emph{disjoint} portions of a program state.
For instance, $x \mapsto \_ \land y \mapsto \_$ asserts that $x$ is a heap cell and $y$ is a heap cell (but they could be the same heap cell), while $x \mapsto \_ * y \mapsto \_$ asserts that $x$ is a heap cell \emph{and separately} $y$ is a heap cell.

Formally, a state $\sigma$ satisfies $P * Q$ if it can be broken up into two disjoint states $\sigma = \sigma_1 \circledcirc \sigma_2$ such that $\sigma_1$ satisfies $P$ and $\sigma_2$ satisfies $Q$ ($\circledcirc$ is a partial operator on states that is effectively disjoint union). When $P$ and $Q$ are formulas denoting heaps, they talk about disjoint regions of the heap, i.e. they do not have any heap addresses in common.
In particular, this means that $x \mapsto \_ * y \mapsto \_$ implies that $x \neq y$, while the formula $x \mapsto y * x \mapsto z$ is unsatisfiable as it is not possible to split a heap into two disjoint portions both of which contain the same address $x$.

In terms of basic concurrent separation logic, propositions mean almost the same as in traditional separation logic; however, we denote ownership by whichever \emph{thread} (execution context) is running the code in question. In the context of the separating conjunction, this means that if a thread \emph{t} can assert $p \mapsto q$, then $t$ knows that no other thread can read or write \emph{p} concurrently, so it can completely ignore the other threads and just reason about \emph{p} as if it were operating in a sequential setting.

\subsubsection{Separating Implication}
The separating implication connective, $\sepimp$ (also known as a \emph{magic wand}), asserts that whenever a fresh heap satisfies a property, its composition with the current heap satisfies another property. This is most useful when a piece of code mutates memory locally, but we want to state some property of the entire memory \cite{spl2}. 

Formally, a state $\sigma$ satisfies $P \sepimp Q$ if for every state $\sigma_1$ that is disjoint from $\sigma$ and that satisfies $P$, the combined state $\sigma \circledcirc \sigma_1$ satisfies $Q$. The best way to understand $\sepimp$ is to think of $*$ and $\sepimp$ as separation logic analogues of $\land$ and $\Longrightarrow$ respectively from first-order logic. For instance, $P * Q$ means you have both $P$ \emph{and} $Q$ (and that they are disjoint). Similarly, $P \sepimp Q$ describes a state such that if you conjoin it with a state satisfying $P$, then you get a state satisfying $Q$.
This property, $P * (P \sepimp Q) \vdash Q$, is the SL analogue of \emph{modus ponens} in first-order logic $P \land (P \Longrightarrow Q) \vdash Q$.

\newpage
\section{Iris Base Logic}\label{base-logic}
Iris is a higher-order concurrent separation logic framework. It defines a program logic for a generic language specified in terms of its expressions, values, and operational semantics; it also provides a weakest-precondition based program logic. Weakest-precondition based program logic stems from Dijkstra's weakest-precondition calculus and is a technique for proving properties of imperative programs.

In this guide we will not provide the grammar and propositions of separation logic because we will show the in-depth discussion of these topics specifically for Iris in this section. We will also refer to Hoare logic specifically in relation to separation logic and Iris. However, it is important to note that Hoare logic is a more general concept and can be used without separation logic for more information refer to \MYhref{http://web.stanford.edu/class/cs357/hoare69.pdf}{Hoare's original paper}. 

\subsection{Propositions}

Iris propositions describe the \emph{resources} owned by a thread as was described above in Section \ref{sep-con}. We will focus on the simple case where propositions describe concrete program states in the form of subsets of the heap. We defer the discussion of advanced resources to more advanced readings\footnote{\MYhref{https://iris-project.org/tutorial-pdfs/iris-lecture-notes.pdf}{Lecture Notes on Iris: Higher-Order Concurrent Separation Logic} provides an in-depth look into many Iris features, including intuition for Iris propositions}.

\begin{figure}[H]
  \begin{align}
    P, Q, R
    &
      := TRUE \mid FALSE \mid P \land Q \mid P \lor Q \mid P \Rightarrow Q  \\
    & \quad
      \mid \exists x.\; P \mid \forall x.\; P \\
    & \quad
      \mid x \mapsto v \mid P * Q \mid P \sepimp Q \\
    & \quad
      \mid \boxed{P}^N  \mid \triangleright P \mid \dboxed{a}^{\gamma}\\
    & \quad 
      \mid \lbrace P \rbrace e \lbrace v.Q \rbrace \mid \langle x. P\rangle {e} \langle v. Q \rangle
      \mid 
  \end{align}
  \caption{The grammar of Iris propositions used in this user guide.}
  \label{fig-grammar-propositions}
\end{figure}\label{fig-1}

The grammar of the subset of Iris propositions that we use throughout the user guide is shown in Figure \ref{fig-1} and includes the following constructs:
\begin{itemize}
\item The first line consists of standard propositional constructs: the propositions $ TRUE$ and $ FALSE$, conjunction, disjunction, and implication.
\item The second line introduces quantification. What makes Iris a \emph{higher-order} logic is that universal and existential quantifiers can range over any type. Hence, $x$ can range over any type, including that of propositions and (higher-order) predicates.
\item We have already seen the points-to proposition, separating conjunction, and separating implication on the third line. 
\item The fourth line contains the invariant proposition $\boxed{P}^N$, explained in \ref{invariants}, the later modality $\triangleright P$, explained in \ref{iris-modals}, and the ghost state proposition $\dboxed{a}^{\gamma}$, explained in \ref{ghost-state}.  
\item Finally, the last line contains the Hoare and atomic triples which will be explained in \ref{hoare}.

\end{itemize}

\subsection{Resource Algebra}
What is a resource? The Iris base logic does not answer this question by fixing a particular set of resources. Instead, the set of resources is kept general and is up to the user of the logic to make a suitable choice. All the logic demands is that the set of resources forms a unital resource algebra (RA)\footnote{Iris actually uses \emph{cameras} as the structure underlying resources, but since we do not use higher-order resources or states which can embed propositions in this guide, we restrict our attention to resource algebras, a stronger, but simpler, structure.} \cite{esop}. We see the formal definition of an RA in Figure \ref{fig-resource-algebra}. 

\begin{figure}[H]
  A \emph{resource algebra} is a tuple
  $(M, \mval : M \rightarrow Prop, (\cdot): M \times M \rightarrow M, \epsilon \in M)$ satisfying:
  \begin{align*}
    \forall a, b, c. \hspace*{2mm} (a \cdot b) \cdot c = a \cdot (b \cdot c) \hspace*{30mm}
     \scriptsize{RA-ASSOC}\\
    \forall a, b. \hspace*{2mm} a \cdot b = b \cdot a \hspace*{30mm}\scriptsize{RA-COMM} \\
    \forall a. \hspace*{2mm} \epsilon \cdot a = a \hspace*{40mm}\scriptsize{RA-ID} \\
    \mval (\epsilon) \hspace*{20mm} \scriptsize{RA-VALID-ID}\\
    \forall a, b. \hspace*{2mm} \mval(a \cdot b)  \Rightarrow \mval(a)  \hspace*{20mm}\scriptsize{RA-VALID-OP}
  \end{align*}
  \caption{The definition of a (unital) resource algebra (RA).}
  \label{fig-resource-algebra}
\end{figure}

Formally, a resource algebra (RA) consists of a set $M$, a validity predicate $\mval$(-), and a binary operation($\cdot$) :$M \times M \rightarrow M$ that satisfy the axioms in Figure \ref{fig-resource-algebra} (Prop is the type of propositions of the meta-logic (e.g., Coq)). RAs are a generalization of the partial commutative monoid (PCM) algebra commonly used by separation logics. In Figure \ref{fig-resource-algebra}, we see the resources owned by different threads can be composed using the $\cdot$ operator, composition of ownership is associative and commutative (reflecting the associative and commutative semantics of parallel composition), and combinations of ownership that do not make sense are ruled out by validity (when multiple threads claim to have ownership of an exclusive resource), specifically a subset, $\mval$, of \emph{valid} elements. This take on partiality will be necessary when we define higher-order ghost states that are needed for modeling invariants which we will discuss in Section \ref{ghost-state} (this is because the composition of all resources at a ghost location must always be valid). This is merely a brief overview of RAs to introduce you to their nature, we encourage readers to delve more into the specifics after grasping the basics first.

\subsection{Invariants}\label{invariants}
We need to  reason compositionally about shared state. Threads take turns interacting with the shared state (reading/writing, etc.). However, how do we verify one thread at a time even though other threads may interfere with, or mutate, the shared state in between each step of computation in the thread we are verifying? Concurrent program logic accounts for such interference via \emph{invariants}. An invariant is a property that holds on some piece of shared state at all times: each thread accessing the state may assume the invariant holds before each step of its computation, but it must also ensure that it continues to hold after each step \cite{invariants-1}.

In Iris, an invariant is a proposition of the form $\boxed{P}^N$ where \emph{P} is an arbitrary Iris proposition. Intuitively, an invariant is a property that, once established, will remain true forever. Therefore, it is a duplicable resource and can be freely shared with any thread. In order to ensure an invariant remains true once it has been established, Iris's proof rules for invariants impose restrictions on how the resources contained in the invariant can be accessed and manipulated. A thread can \emph{open} an invariant $\boxed{P}^N$ to gain ownership of the contained resources \emph{P}. These resources can then be used in the proof of a single atomic step of the thread's execution. After the thread has performed an atomic step with an open invariant, the invariant must be closed, which amounts to proving that \emph{P} has been reestablished. Otherwise, the proof will not succeed.

In our denotation of an invariant, $\boxed{P}^N$, the \emph{N} refers to the namespace of the invariant. A namespace is the part of the mechanism used in Iris to keep track of invariants that are currently open and need to be closed before the next atomic step. This ensures we avoid issues of re-entrancy in case of nested invariants which would lead to logical inconsistencies.

In Section \ref{case-work} we explain a "parallel increment" example. In this example we define the body of an invariant and explain how we open and close that invariant. Through this process we will see the main tactics used for this type of procedure.

\subsection{Modalities}\label{modal}
Modalities are expressions that qualify assertions about the truth of statements. Iris extends traditional separation logic with a handful of modalities - such as the persistence modality ($\square$), the step-indexed "later" modality ($\triangleright$), and the (basic) update modality ($\Rrightarrow$ \emph{P}). Iris proof mode provides tactical support for reasoning conveniently about these modalities, some of which are introduced in Section \ref{iris-modals}. 
\begin{itemize}
    \item Persistence modality : denoted by $\square P$, asserts that \emph{P} holds without asserting any exclusive ownership. An assumption, $\square P$, can be used arbitrarily often, it cannot be "used up". For example, $P \sepimp Q$ is a linear implication that can be applied only once. However, $\square (P \sepimp Q)$ can be applied arbitrarily often \cite{iris-ground-up}. Note that this can be used to encode Hoare triples (explained in Section \ref{hoare}). An assertion \emph{P} is persistent if proofs of \emph{P} can never assert exclusive ownership, this means that these propositions can be freely duplicated, the usual restriction of resources being usable only once does not apply to them.
    \item Later modality : denoted by $\triangleright P$, asserts that \emph{P} holds at the next step-index or "one step later". Iris assertions can serve as an invariant, so one can define invariants that refer to other invariants or even themselves. This introduces a potential circularity. The notion of steps using a later modality introduces step-indexing which allows us to break this circularity. To ensure that invariants are well defined, the shared resource backing $\boxed{P}^N$ need satisfy only $\triangleright P$. Therefore, opening an invariant grants ownership of the shared resource one step later and closing the invariant reestablishes the shared resource one step later \cite{invariants-1}. Thus we see the later modality is needed to ensure soundness in complex cases when Hoare triples or invariants themselves are stored inside invariants. If we do not store Hoare triples or invariants inside invariants, we can generally ignore the later modality, which we will do throughout this guide \cite{iris-ground-up}. 
    \item (Basic) Update modality : denoted by $\Rrightarrow P$, allows you to talk about what you could own after performing an update to what you do own \cite{esop}. It reflects frame-preserving updates into the logic, in the sense that $\Rrightarrow P$ asserts ownership of resources that can be updated to resources satisfying \emph{P}. A frame-preserving update is an update to a resource. If we think of frames as being the resources owned be other threads, then a frame-preserving update is guaranteed not to invalidate the resources of concurrently-running threads. There is another modality, the fancy update modality or view shift, that has a similar, but more advanced purpose which will not be covered in this guide but is covered more in \MYhref{https://iris-project.org/pdfs/2017-esop-iris3-final.pdf}{The Essence of
Higher-Order Concurrent Separation Logic} paper.
\end{itemize}

\label{ghost-state}
\subsection{Ghost State}
\emph{Ghost State}, originally called auxiliary variables, is a formal technique where the prover adds state (variables or resources) to a program that captures knowledge about the history of a computation not present in the state of the original program in order to verify it. Ghost state was originally proposed as a way to abstractly characterize some knowledge about the history of a computation that is essential to verifying it. It is used for modularly describing a thread's knowledge about some shared state as well as the rights it has to modify it. Ghost state is a purely logical construct, introduced solely for the purpose of verification, and is not a part of the program syntax. Since in Iris, ghost state is purely logical and ghost commands are represented as proof rules, we know that the added ghost state and the ghost commands that modify it, have no effect on the run-time behavior of the program. The proof of this augmented program can be transformed into a proof of the program with all ghost states removed (i.e., the original program)\footnote{Iris does not distinguish between real and ghost states, all states are considered to be ghost states.}.

\subsection{Hoare Triples and Atomic Triples}\label{hoare}
Hoare Triples are assertions of the form  $\lbrace P \rbrace e \lbrace v.Q \rbrace$, where $P$ and $Q$ are formulas.
$\lbrace P \rbrace e \lbrace v.Q \rbrace$ is true if for every state $\sigma$ that satisfies $P$ we have that (1) the program $e$ does not reach an error state when run from $\sigma$ (for example, by trying to read unallocated memory), and (2) that if $e$ terminates then it returns some value $v$ and the new state is some $\sigma'$ that satisfies $Q$.
We call $P$ the \emph{precondition} and $Q$ the \emph{postcondition} of $e$, and we write $\lbrace P \rbrace e \lbrace Q \rbrace$ in the case where $Q$ does not mention the return value $v$. Let us look at an example of a Hoare Triple:
\[
   \lbrace X = 1\rbrace X:=X+2 \lbrace X = 3\rbrace
\]
Here $P$ is the condition that the value of X is 1, $Q$ is the condition that the value of X is 3, and $e$ is the assignment command $X:=X+2$ i.e. ‘X becomes X+2’. We know this is a valid Hoare Triple as the command X := X + 2 transforms a state in which X = 1 to a state in which X = 3 \cite{hoare-1}.

Recall the mention of weakest-precondition from the beginning of this section and let us break down what this means using Hoare Logic. Given a precondition $P$, a command $e$, and a postcondition $Q$ such that $\lbrace P \rbrace e \lbrace Q \rbrace$, we want to find the unique $P$ that is the weakest precondition for $e$ and $Q$. This means that if $e$ is a command and  $Q$ is an assertion about states, then the weakest precondition for $e$ with respect to $Q$ is an assertion that is true for precisely those initial states from which $e$ must terminate and executing $e$ must produce a state that satisfies $Q$. Let us see this in an example where $e$ is $X:=X+1$ and $Q$ is $(X>0)$. One valid precondition would be $(X>0)$ as we see the following Hoare statement is valid:
\[
   \lbrace X>0\rbrace X:=X+1 \lbrace X>0\rbrace
\]
Another valid precondition would be $(X>-1)$ as the following Hoare statement is also valid:
\[
   \lbrace X>-1\rbrace X:=X+1 \lbrace X>0\rbrace
\]
We can see that $(X>-1)$ is weaker than $(X>0)$ since $(X>0)\Rightarrow (X>-1)$ and in fact, $(X>-1)$ is our weakest precondition \cite{wp-1}. 

Logical atomicity is a useful concept that is used to reason about concurrent structures. It is used to specify programs that execute in multiple atomic steps but whose effect appears to take place in a single point in time. This concept will be useful for specifying concurrent structures in a way that can be used to verify concurrent client programs.

We can specify the concurrent behavior of programs using \emph{atomic triples}. An atomic triple $\langle x. P\rangle {e} \langle v. Q \rangle$ is made up of the pre-condition $P$, return value $v$, post-condition $Q$, and a program $e$. Such a triple means that $e$, if it terminates, despite executing in potentially many atomic steps, appears to operate atomically on the shared state and transforms it from a state satisfying $P$ to one satisfying $Q$. This means there is a single physical step during the execution of $e$ when the shared state is transformed from $P$ to $Q$.

For the sake of this guide, we briefly introduce the concept of atomic triples before diving into proof rules and propositions. We will introduce proof rules for Hoare and Atomic Triples in Section \ref{sep-formulas} so we can use them to reason about programs and examples in this guide.

\subsection{Proof Rules}\label{sep-formulas}
Before we can prove any specifications in separation logic, we must introduce some proof rules for Hoare Triples (Figure 2.1). A proof rule, or inference rule, consists of two parts separated by a horizontal line: the part above the line contains one or more premises, and the part below the line contains the conclusion. A rule with no premises is called an \emph{axiom}, and in this case we omit the horizontal line. Some rules are bi-directional: the premise implies the conclusion, but the conclusion also implies the premise. These rules are denoted with a double horizontal line. We will explain some of the proof rules in Figure \ref{fig-hoare-proof-rules} as we discuss an example below. For more information on the proof rules, reference the \MYhref{https://iris-project.org/tutorial-pdfs/iris-lecture-notes.pdf}{Lecture Notes on Iris: Higher-Order Concurrent Separation Logic}. 
\begin{figure}[H]
  \begin{mathpar}
    \inferH{hoare-ret}
    {}
    {\hoareTriple{True}{w}{v. v = w}}

    \inferH{hoare-false}
    {}
    {\hoareTriple{False}{e}{v. P}}

    \inferH{hoare-alloc}
    {}
    {\hoareTriple{True}{\refCommand(v)}{\ell. \ell \mapsto v}}

    \inferH{hoare-load}
    {}
    {\hoareTriple{\ell \mapsto v}{!v}{w. \ell \mapsto v \; *  \; w = v}}

    \inferH{hoare-store}
    {}
    {\hoareTriple{\ell \mapsto v}{\ell \gets w}{\ell \mapsto w}}
    
  \end{mathpar}

  \hrule

  \begin{mathpar}
    \inferH{hoare-lam}
    {\hoareTriple{P}{e[x \goesto v]}{w. Q}}
    {\hoareTriple{P}{\lambdaFn{x}{e} \; v}{w. Q}}

    \inferH{hoare-fork}
    {\hoareTriple{P}{e}{True}}
    {\hoareTriple{P}{\forkCommand\lbrace e\rbrace}{True}}

    \inferH{hoare-let}
    {\hoareTriple{P}{e_1}{w. R} \\
      \forall w. \hoareTriple{R}{e_2[x \goesto w]}{v. Q}}
    {\hoareTriple{P}{\letCommand{x}{e_1}{e_2}}{v. Q}}
  \end{mathpar}

  \hrule

  \begin{mathpar}
    \inferH{hoare-csq}
    {P \Rrightarrow  P' \\
      \hoareTriple{P'}{e}{v.Q'} \\
      \forall v. Q' \Rrightarrow Q}
    {\hoareTriple{P}{e}{v.Q}}

    \inferH{hoare-frame}
    {\hoareTriple{P}{e}{v. Q}}
    {\hoareTriple{P * R}{e}{v. Q * R}}

    \inferHB{hoare-disj}
    {\hoareTriple{P}{e}{v.R} \land \hoareTriple{Q}{e}{v.R}}
    {\hoareTriple{P \lor Q}{e}{v.R}}

    \inferHB{hoare-exist}
    {\forall x. \hoareTriple{P}{e}{v.Q}}
    {\hoareTriple{\exists x. P}{e}{v.Q}}
  \end{mathpar}

  \caption{Some of the proof rules for establishing Hoare triples. We write $e[x \goesto v]$ to denote the expression $e$ after substituting all occurrences of the variable $x$ with the term $\val$.}
  \label{fig-hoare-proof-rules}
\end{figure}

Recall our discussion of backward reasoning, we can apply this process to solving a proof tree. Like mentioned before, we will start with a specification that we want to prove, find an inference rule that matches the structure of the goal, and apply it. This replaces the goal with the premises of the rule. 

Let us take a counter example to better understand how we prove a program in this style. Take the following program expression which reads the value stored at heap location $x$ into a variable ($v \;=\; !x$) and then writes $v + 1$ back into the location:
\[
   e_{inc} := \letCommand{v}{!x}{x \gets (v + 1)}
\]
Essentially, this means $e_{inc}$ is a program that increments the value stored at the heap location $x$. Before we can prove anything, we need to use Hoare Triples to write the specifications of our program $e_{inc}$ as follows: 
\[
  \hoareTriple{x \mapsto n}{e_{inc}}{x \mapsto n + 1}
\]

We see that here the precondition uses a points-to predicate $x \mapsto n$ to assert the program state contains a heap cell at address $x$ containing value $n$ and the postcondition uses a points-to predicate to assert the program state contains a heap cell at address $x$ containing value $n + 1$. In simple words, this Hoare Triple tells us that if the program $e_{inc}$ is run on a state that contains a heap cell at address $x$ with value $n$, then it results in a state where the cell $x$ contains $n + 1$.

We can now prove that $e_{inc}$ meets this specification using our proof rules as follows:

\begin{prooftree}
  \footnotesize
  \AxiomC{$hoare-load$}
  \AxiomC{\qquad\qquad\qquad}
  \BinaryInfC{$\hoareTriple{x \mapsto n}{!x}{\Ret\val. x \mapsto n \; * \; v = n}$}
  \AxiomC{}
  \RightLabel{$hoare-store$}
  \UnaryInfC{$\hoareTriple{x \mapsto n}{x \gets (v + 1)}{x \mapsto v + 1}$}
  \RightLabel{$hoare-frame$}
  \UnaryInfC{$\hoareTriple{x \mapsto n \; * \; v = n}{x \gets (v + 1)}{x \mapsto v + 1 \; * \; v = n}$}
  \RightLabel{$hoare-csq$}
  \UnaryInfC{$\forall v. \hoareTriple{x \mapsto n \; * \; v = n}{x \gets (v + 1)}{x \mapsto n + 1}$}
  \RightLabel{$hoare-let$}
  \BinaryInfC{$\hoareTriple{x \mapsto n}{\letCommand{v}{!x}{x \gets (v + 1)}}{x \mapsto n + 1}$}
\end{prooftree}

When explaining a proof using backward reasoning, we use a proof tree. We read \emph{proof trees} starting from the goal at the bottom or the root of the tree and work our way up.
Each horizontal line represents the use of an inference rule to transform the current goal into one or more goals. As we apply more inference rules, our goal hopefully becomes simpler. 

We will briefly explain our proof tree for $e_{inc}$ above, although certain details will be emitted for readability for beginners. In this proof tree, we see the first rule we use is $ HOARE-LET$, which works on let-bindings and breaks up the proof. Here we see it splits the proof of the bound expression $!x$ (on the left), and the proof of the let-body $x \gets (v+ 1)$ (on the right). It is important to note that proof rule $ HOARE-LET$ requires the user to come up with an intermediate proposition $R$ that describes the states of the program after evaluating the bound expression $e_1$ but before evaluating the let-body $e_2$.  In this particular example since $e_1$ is a dereference, our only option is to use the rule $ HOARE-LOAD$, allowing us to solve the left sub-tree. Now let us consider the right sub-tree. Here our only option is to use $ HOARE-STORE$, but the current proof goal does not quite fit. Specifically, $HOARE-STORE$ will give us the postcondition $x \mapsto v + 1$ while we need the postcondition $x \mapsto n + 1$. In order to infer this, we need the fact that $v = n$, which is something that we know in the precondition of the current goal. If we can transfer the fact $v = n$ from the precondition to the postcondition of the program fragment $x \gets (v + 1)$, then we can use $ HOARE-CSQ$ to complete the proof tree. 

Since we can think of propositions in separation logic as resources, let us think about the predicate $x \mapsto n$ as a resource since it describes a heap cell or a part of the program state. The \emph{frame rule}, $ HOARE-FRAME$, allows us to \emph{frame} resources around a program fragment. This means we can carry resources that are not needed by the program fragment from its precondition to its postcondition. In the case of our proof tree, the program fragment is $x \gets (v + 1)$, and as this does not modify $v$ or $n$, we frame the resource $v = n$, allowing us to use $ HOARE-STORE$ to finish the proof.

As can be seen from this example, the structure of the program expression will often determine the proof rule we must use, as well as any intermediate proposition that links the first goal to the next one.

\newpage
\section{Iris in Action}
\label{iris-discussion}
 For the purpose of this guide, we refer to Iris implemented and verified in the Coq proof assistant. This section is dedicated to the different pieces of Iris to give insight into the highly modular framework and how to use Iris in the Coq proof assistant \cite{mit}. Earlier sections introduce Iris base logic and the core logic that all of Iris is built upon. Later, we will also present further examples.

\subsection{HeapLang}\label{heaplang}
HeapLang is the standard example language for Iris. While it is not the only language we can reason about with Iris, it is a reasonable language for simple examples. HeapLang is a lambda-calculus programming language with mutable state recursive functions, general higher-order references and fork-style concurrency (the fork construct can be used to spawn new threads). Integers, Booleans, units, heap locations, and (binary) sums and products are built-in primitives \cite{heap-lang-1, coq2}. It has syntax for Hoare triples and tactics for relatively easy program proofs of weakest preconditions. Here we will explain the basics of using HeapLang in a Coq proof assistant; however, for a more thorough list of tactics, please refer to the \MYhref{https://gitlab.mpi-sws.org/iris/iris/-/blob/master/docs/heap_lang.md}{HeapLang documentation}. An explanation on how to install HeapLang will be addressed in Section \ref{install-iris}. 

\subsubsection{Tactics}
HeapLang comes with tactics that facilitate stepping through HeapLang programs as part of proving a weakest precondition. Note that here, \# is used as a short-hand to turn a basic literal (an integer, a boolean literal, etc.) into a value. Since values coerce to expressions, \# is used whenever a Coq value needs to be placed into a HeapLang term. Now let us look at some important tactics that take one or more pure program steps: 
\begin{itemize}
    \item $\textcolor{blue}{\textbf{wp\_pure}}$ : performs one pure reduction step. Pure steps include beta reduction, projections, constructors, as well as unary and binary arithmetic operators.
    \item $\textcolor{blue}{\textbf{wp\_pures}}$ : performs as many pure reductions steps as possible. This tactic will not reduce lambdas that are hidden behind a definition. If the computation reaches a value, the WP will be entirely removed and the postcondition becomes the new goal.
    \item $\textcolor{blue}{\textbf{wp\_bind pat}}$ : apply the bind rule to "focus" the term matching pat. This is useful when accessing invariants, which is possible only when the WP in the goal is for a single, atomic operation. wp\_bind can be used to bring the goal into the right shape.
    \item $\textcolor{blue}{\textbf{wp\_apply proof\_mode\_term}}$ : apply a lemma whose conclusion is a WP, automatically applying wp\_bind as needed. 
    \item $\textcolor{blue}{\textbf{wp\_load}}$ : reduces a load operation and automatically will find the points-to assertion in the spatial context.
    \item $\textcolor{blue}{\textbf{wp\_store}}$ : reduces a store operation and automatically will find the points-to assertion in the spatial context. 
    \item $\textcolor{blue}{\textbf{wp\_op}}$ : reduces unary, binary and other arithmetic operators. 
\end{itemize}
All of the tactics of HeapLang assume that our current goal is of the form "WP e @ E $\lbrace \lbrace$ Q $\rbrace \rbrace$" (where WP stands for "weakest precondition"). Let us take a look at one of our examples from Section \ref{sepLogic} to see how some of these tactics work. Take the Hoare triple: $\lbrace X = 1\rbrace X:=X+2 \lbrace X = 3\rbrace$. We won't go into the implementation specifics in this section, but we will show how these tactics work once we have our "WP" format. As we can see in Figure \ref{1,3}. Here we will be looking at our program "inc\_x" which stands for the function, X := X + 2. Since we see a goal in this format, we know to use HeapLang tactics. 

\begin{figure}[H]
\centering
\begin{lstlisting}[mathescape=true]
Definition wp : expr := 
$\lambda$: "l", let: "n" := !"l" in
        "l" <- "n" + #2.

Lemma wp_example ($\ell$ : loc) (n : Z):
{{{ $\ell$ $\mapsto$ #1 }}} wp #$\ell$ {{{RET #(); $\ell$ $\mapsto$ #3}}}.

Proof.
iIntros ($\Phi$) "Hpt H$\Phi$.
wp_pures. wp_load. wp_store. iModIntro.
iApply "H$\Phi$". iFrame.
Qed.
\end{lstlisting}
\end{figure}

\begin{figure}[H]
  \includegraphics[width = 150mm]{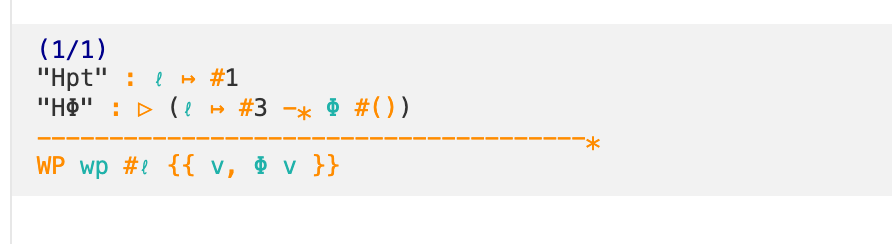}
  \centering
  \caption{Our current goal is in the format "WP e @ E $\lbrace \lbrace$ Q $\rbrace \rbrace$".}
  \label{1,3}
\end{figure}

\begin{figure}[H]
  \includegraphics[width = 150mm]{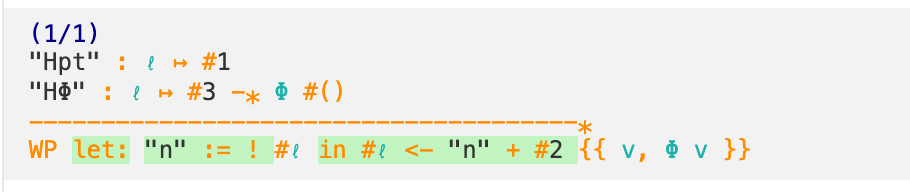}
  \centering
  \caption{The tactic wp\_pures reduces as many steps as possible.}
  \label{1,3,pures}
\end{figure}

\noindent For example, we will use the tactic $\textcolor{blue}{\textbf{wp\_pures}}$, one of the most common HeapLang tactics, to perform as many pure reduction steps as possible. This could potentially be the last HeapLang tactic we need. If our computation reaches a value, the WP will be entirely removed from the goal. However, we see in Figure \ref{1,3,pures} this is not the case. 

\begin{figure}[H]
  \includegraphics[width = 150mm]{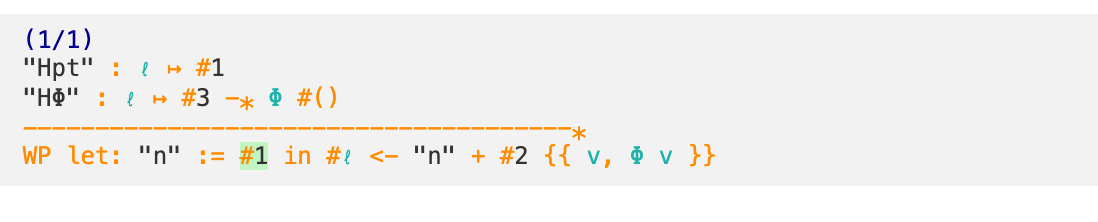}
  \centering
  \caption{The tactic wp\_load will reduce the load operation and introduce \#1 into our goal.}
  \label{1,3,load}
\end{figure}

\noindent We know we must still use another HeapLang tactic before we can move on from this portion of the proof. We see in Figure \ref{1,3,pures} we see that in our \emph{let} statement "n" := ! \#$\ell$. We want to reduce this load operation by using the tactic $\textcolor{blue}{\textbf{wp\_load}}$. This will take the "1" that we assumed in our precondition and introduce it into our goal. The last step we see after this is to somehow combine our 1 and 2 in the goal in order to get the 3 that we want. We can use the tactic $\textcolor{blue}{\textbf{wp\_store}}$ to automatically find the points-to assertion in our goal that is "$\# \ell$ <- "n" + \#2" and combine it with "n" := \#1. This will then get stored in our hypothesis "Hpt" as seen in Figure \ref{1,3,store}. 

\begin{figure}[H]
  \includegraphics[width = 150mm]{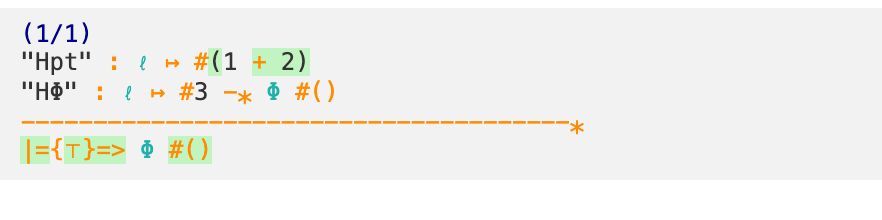}
  \centering
  \caption{The tactic wp\_store will automatically find the points-to assertion in our goal that is "$\# \ell$ <- "n" + \#2" and combine it with "n" := \#1.}
  \label{1,3,store}
\end{figure}

\noindent Since we no longer see a "WP" in our current goal we know that we have finished the weakest precondition portion of our proof and will no longer be using our HeapLang tactics. Although we show all of the code needed for this example below, Section \ref{case-work} will provide further details into the implementation of similar proofs and a deep explanation into tactics required to start and finish such a proof.

\subsection{Iris Proof Mode}\label{IPM}
The Iris Proof Mode (IPM) is an embedded domain specific language (DSL) that is used for interactive proofs in separation logic. It works for a large variety of separation logics. The IPM is actually mostly independent of Iris as the implementation uses only tactics and typeclasses. We will introduce many of the important tactics IPM provides here, however, for a more in-depth guide with more tactics, refer to the \MYhref{https://gitlab.mpi-sws.org/iris/iris/blob/master/docs/proof_mode.md}{Iris tactic documentation}. These tactics will be used throughout this guide to prove various examples.

The example we will utilize in this section to describe some of the tactics we introduce will be proving transitivity of x, y, and z. The whole code for this example is provided in Figure \ref{trans-code} but we will continue to break down individual tactics throughout this section. The tactics we do not provide an example for here will be explained in the case work at the end of this guide. There we will also explain the different parts of a proof and how to bring them all together. Below we introduce the full code for our transitivity example, but specific proof views and tactics will be explained as we progress through this section. Note that the tactics $\textcolor{blue}{\textbf{rewrite}}$, $\textcolor{blue}{\textbf{subst}}$, and $\textcolor{blue}{\textbf{done}}$ are Coq tactics so will not be explained here, for further information on their purpose, refer to Section \ref{coq}. 

\begin{figure}[H]
\centering
\begin{lstlisting}[mathescape=true]
Definition trans : expr := 
$\lambda$: "x", 
  let: "y" := "x" in
  let: "z" := "y" in 
  "x" = "y";;
  "y" = "z";; 
  #().

Lemma transitivity (x y z: loc):
  {{{ $\ulcorner$x = y$\urcorner$ $*$ $\ulcorner$y = z$\urcorner$ $*$ $\ulcorner$x = x$\urcorner$}}} trans #x {{{RET #(); $\ulcorner$x = z$\urcorner$}}}.
  
Proof.
  iIntros ($\Phi$) "Hxy H$\Phi$".
  iRename "Hxy" into "Hx".
  iDestruct "Hx" as "(Hxy & Hyz & Hxx)".
  iClear "Hxx".
  iDestruct "Hxy" as %Hxy.
  iDestruct "Hyz" as %Hyz.
  wp_pures. iModIntro.
  iApply "H$\Phi$".
  iAssert ($\ulcorner$x = z$\urcorner$)%I as %yz.
  {
    iPureIntro. rewrite Hyz in Hxy. done.
  } subst z. 
  iAssert ($\ulcorner$x = y$\urcorner$)%I as %xy.
  {
    iPureIntro. done.
  } subst y. iFrame.
Qed.
\end{lstlisting}
\caption{Proving the transitivity of x, y, and z}
\label{trans-code}
\end{figure}

\subsubsection{Basic Terms}
Before we look at our example and tactics, let us consider Figure \ref{example-proof-setup} as an example proof view in IPM. The IPM embeds a separation logic context within the Coq goal. This means we have the Coq context above the solid line and the IPM context below the solid line. Inside the IPM context we have a persistent context (intuitionistic separation logic hypotheses) and a spatial context (spatial separation logic hypotheses). The persistent context is separated by - - - -$\square$ and is composed of facts that are duplicable and do not go away when split, they can be used any number of times. The spatial context is separated by - - - -$*$ and is composed of ordinary spatial premises. More explanation into these different contexts and the tactics that must be used for each one are explained in the sections below. To follow up this discussion, Section \ref{case-work} contains in-depth examples of many of these tactics and how they might effect the context of a proof. 

\begin{figure}[H]
    $\scriptsize{\overrightarrow{x} : \overrightarrow{\Phi}}$ \hspace*{2mm} Variables and pure Coq hypotheses \newline
\rule{75mm}{.5mm} \newline
$\scriptsize{\pi_2}$ \hspace*{2mm} Intuitionistic separation logic hypotheses \newline
- - - - - - - - - - - - - - - - - - - - - - - - - - - - - - - -$\square$ \newline
$\scriptsize{\pi_1}$ \hspace*{2mm} Spatial separation logic hypotheses \newline
- - - - - - - - - - - - - - - - - - - - - - - - - - - - - - - -$*$ \newline
$\scriptsize{R}$ \hspace*{2mm} Separation logic goal
    \caption{Example proof set-up in Iris Proof Mode that corresponds to proving: \newline $\Pi_2 \sepimp \Pi_1 \sepimp R$}
    \label{example-proof-setup}
\end{figure}

\subsubsection{Introduction patterns (ipat)}
Introduction patterns are used to perform introductions and eliminations of multiple connectives. This means we will either add a hypothesis or remove a hypothesis from our proof state. Some of the introduction patterns that the proof mode supports are:
\begin{itemize}
    \item $\textcolor{blue}{\textbf{H}}$ : creates a hypothesis named H. In the transitivity example presented above, we do not create a hypothesis "H" but rather introduce hypotheses "Hxy" and "H$\Phi$". We can later use these hypotheses to further break down our goal. Note that naming of these hypotheses is up to the user. 
\begin{figure}[H]
\centering
  \includegraphics[width = 150mm]{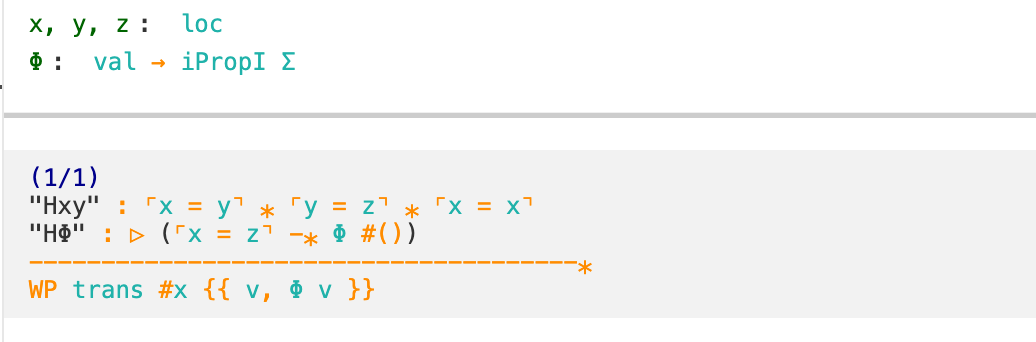}
  \caption{The result of using the tactic iIntros ($\Phi$) "Hxy H$\Phi$" to introduce hypothesis "Hxy" and "H$\Phi$ into the spatial context of our proof state.}
\end{figure}
    \item $\textcolor{blue}{\textbf{?}}$ : creates an anonymous hypothesis (unnamed hypothesis).
    \item $\textcolor{blue}{\textbf{\_}}$ : clears the hypothesis out of the proof view.
    \item $\textcolor{blue}{\textbf{\$}}$ : frames the hypothesis in the goal of the proof view.
    \item $\textcolor{blue}{\textbf{$[$ipat1 ipat2$]$}}$ : (separating) conjunction elimination (separating conjunction explained in Section \ref{sep-con}). In order to destruct conjunctions of the form $P \land Q$, either the proposition P or Q must be persistent \emph{or} either ipat1 or ipat2 should be \_ which allows the result of one of the conjuncts to be thrown away (hence it will disappear from the proof view). 
\end{itemize}
Note that in case a branch of ipat starts with \#, it causes the hypothesis to be moved to the intuitionistic context and if it starts with \%, it causes the hypothesis to be moved to the pure Coq context. 
\subsubsection{Selection Patterns (selpat)}
Selection patterns are used to select hypotheses in certain tactics such as iFrame and iLöb. Some of the selection patterns that the proof mode supports are:
\begin{itemize}
    \item $\textcolor{blue}{\textbf{H}}$ : select the hypothesis named H.
    \item $\textcolor{blue}{\textbf{\%}}$ : select the entire pure/Coq context. In the transitivity example presented above, we destruct some of our hypotheses into pure/Coq context so that we can use them later in our subproofs. We do this by placing \% in front of the hypothesis name as seen in the tactic: iDestruct "Hxy" as \%Hxy. This places the hypothesis into our Coq context as seen in Figure \ref{coq_trans}. 
\begin{figure}[H]
\centering
  \includegraphics[width = 150mm]{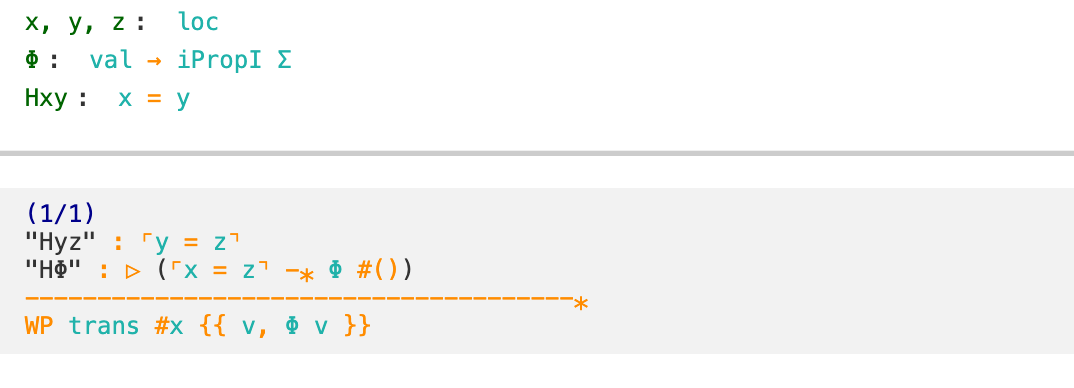}
  \caption{The result of using the tactic iDestruct "Hxy" as \%Hxy to place hypothesis "Hxy" into the Coq context.}
  \label{coq_trans}
\end{figure}
    \item $\textcolor{blue}{\textbf{\#}}$ : select the entire intuitionistic context.
    \item $\textcolor{blue}{_{*}}$ : select the entire spatial context, note that $_{*}$ is the unicode symbol $_{*}$ and NOT the regular asterisk *.
\end{itemize}
\subsubsection{Applying hypotheses and lemmas}
Many of the IPM tactics can take both hypotheses and lemmas and allow one to instantiate universal quantifiers and implications or wands with these hypotheses and lemmas. We consider these proof mode terms (pm\_trm). Some of the tactics in this category include:
\begin{itemize}
    \item $\textcolor{blue}{\textbf{iExact "H"}}$ : this will finish the current goal of the proof if the conclusion matches the hypothesis H.
    \item $\textcolor{blue}{\textbf{iApply pm\_tm}}$ : this will match the conclusion of the current goal against the conclusion of the pm\_trm and generates goals for the premises of pm\_trm. In the transitivity example presented above, we eventually have the goal of "$\phi$\#()" as seen in Figure \ref{apply1_trans}. We see that this matches the right hand side of our hypothesis "H$\Phi$". In order to get rid of this goal we can use the tactic iApply "H$\Phi$" which will apply the right hand side so that we are left with proving the left hand side of the hypothesis as seen in Figure \ref{apply2_trans}. 
\begin{figure}[H]
\centering
  \includegraphics[width = 150mm]{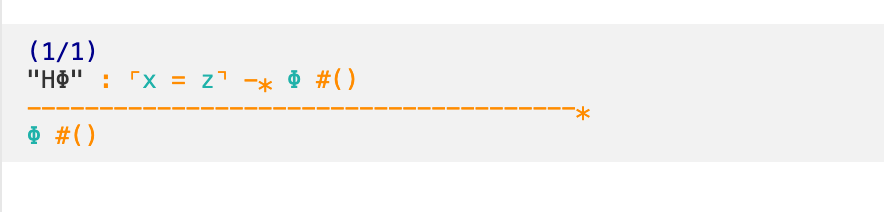}
  \caption{The proof state when we have goal "$\phi$\#()".}
  \label{apply1_trans}
\end{figure}
\begin{figure}[H]
\centering
  \includegraphics[width = 150mm]{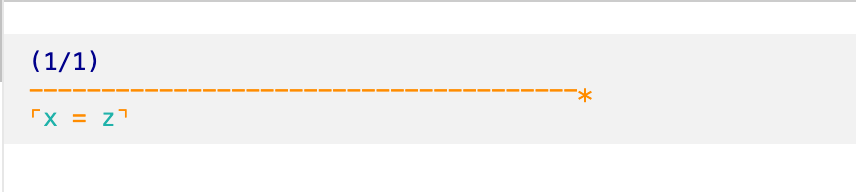}
  \caption{The result of using the tactic iApply "H$\Phi$" which will apply the right hand side of "H$\Phi$" so that we are left with only proving the left hand side of "H$\Phi$".}
  \label{apply2_trans}
\end{figure}

\end{itemize}
\subsubsection{Context management}
Some of the tactics in this category include: 
\begin{itemize}
    \item $\textcolor{blue}{\textbf{iIntros \((x_1 ... x_n)\) "ipat1 ... ipatn"}}$ : introduces universal quantifiers using Coq introduction patterns \(x_1 ... x_n\) and implications/wands using proof mode introduction patterns ipat1 ... ipatn. In the transitivity example presented above, we introduce the hypotheses "Hxy" and "H$\Phi$". This tactic unfolds our triple notation and introduces the resulting hypotheses into our spatial context. 
\begin{figure}[H]
\centering
  \includegraphics[width = 150mm]{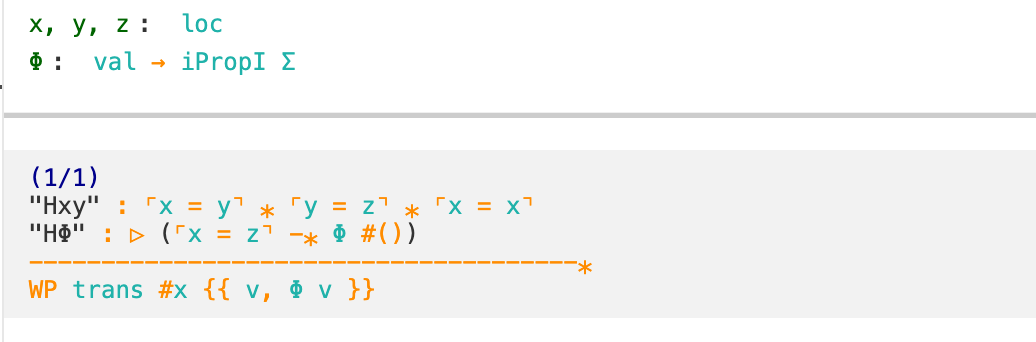}
  \caption{The result of using the tactic iIntros ($\Phi$) "Hxy H$\Phi$" to introduce hypothesis "Hxy" and "H$\Phi$ into the spatial context of our proof state.}
\end{figure}
    \item $\textcolor{blue}{\textbf{iRename "H1" into "H2"}}$ : renames the hypothesis H1 into H2. In the transitivity example presented above, we rename our hypothesis "Hxy" to be "Heverything" because it is a more fitting name. As can be seen in the proof view below, the only thing that changed after our tactic call was the name of our hypothesis. 
\begin{figure}[H]
\centering
  \includegraphics[width = 150mm]{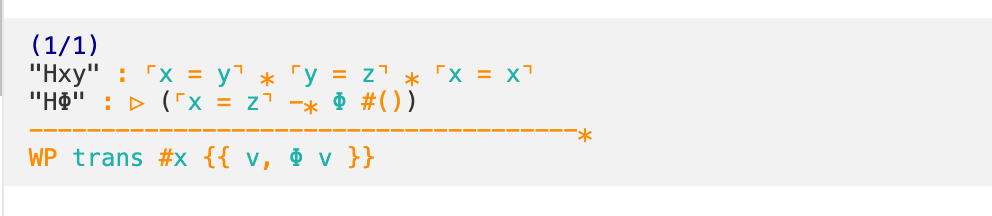}
  \caption{Before using the tactic iRename "Hxy" into "Heverything" we have hypotheses "Hxy" and "H$\Phi$".}
\end{figure}
\begin{figure}[H]
  \includegraphics[width = 150mm]{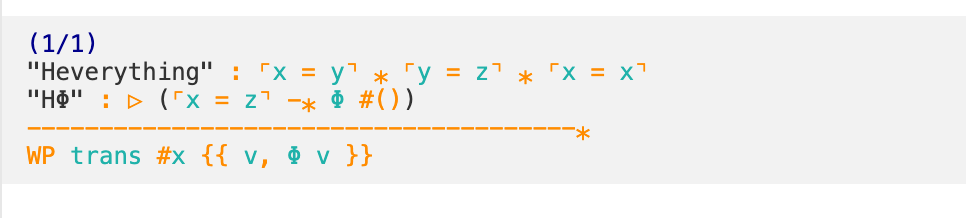}
  \caption{After using the tactic iRename "Hxy" into "Heverything" we have hypotheses "Heverything" and "H$\Phi$".}
\end{figure}
    \item $\textcolor{blue}{\textbf{iClear \((x_1 ... x_n)\) "selpat"}}$ : clear the hypotheses given by the selection pattern selpat and the Coq level hypotheses variables \(x_1 ... x_n\). In the transitivity example presented above, we notice that we have a hypothesis denoting "x=x". This is completely unneccessary for our proof, so we can use the tactic iClear in order to help keep our proof view easily readable. This tactic will merely get rid of the hypothesis/hypotheses you do not need. Note that this is most often used in large proofs when you have too many hypotheses and need to clear some to better visualize the goal at hand. 
\begin{figure}[H]
\centering
  \includegraphics[width = 150mm]{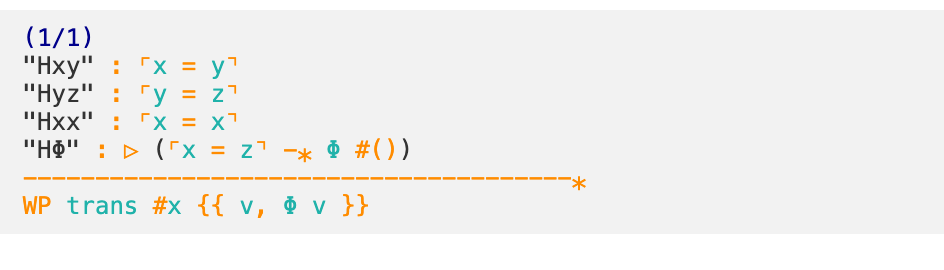}
  \caption{Before using the tactic iClear "Hxx" we have the hypothesis "Hxx".}
\end{figure}
\begin{figure}[H]
  \includegraphics[width = 150mm]{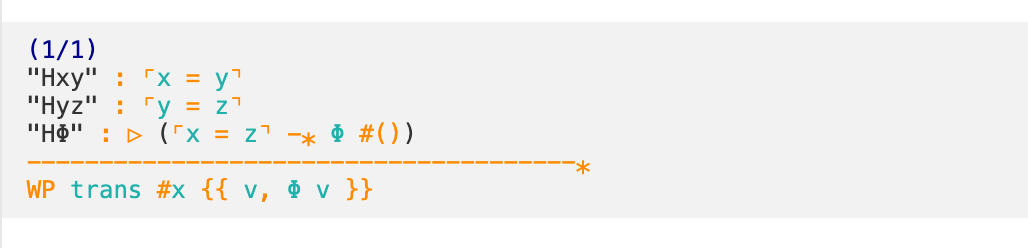}
  \caption{After using the tactic iClear "Hxx" we no longer have the hypothesis "Hxx".}
\end{figure}
    \item $\textcolor{blue}{\textbf{iPoseProof proof\_mode\_term as \((x_1 ... x_n)\) "ipat"}}$ : puts the proof mode term into the context and destructs it using the introduction pattern ipat. This tactic is essentially the same as iDestruct, the difference being that the proof mode term is not thrown away if possible. 
    \item $\textcolor{blue}{\textbf{iAssert (P)\%I with "specialization\_pattern" as "H"}}$ : generates a new subgoal P and adds the hypothesis P to the current goal as H. The specialization pattern specifies which hypotheses will be consumed by proving P. In the transitivity example presented above, we use this tactic multiple times. Let us look at the first case: iAssert ($\ulcorner$x = z$\urcorner$)\%I as \%yz. When we use this tactic, it creates a new subgoal where we will need to prove what we assert. In this case we will prove the subgoal $\ulcorner$x = z$\urcorner$ as seen in Figure \ref{assert_trans}. After we prove this goal, it will be placed in our Coq context as the hypothesis "yz" denoted by \%yz.
\begin{figure}[H]
\centering
  \includegraphics[width = 150mm]{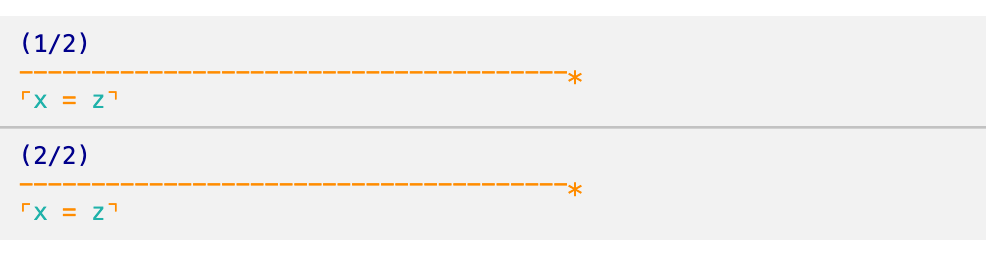}
  \caption{The result of using the tactic iAssert ($\ulcorner$x = z$\urcorner$)\%I as \%yz. This tactic creates a new subgoal where we will need to prove $\ulcorner$x = z$\urcorner$.}
  \label{assert_trans}
\end{figure}
\end{itemize}
\subsubsection{Introduction of logical connectives}
Some of the tactics in this category include:
\begin{itemize}
    \item  $\textcolor{blue}{\textbf{iPureIntro}}$ : turn a pure goal, typically of the form $\ulcorner \varphi \urcorner$, into a Coq goal. Here $\ulcorner \hspace*{2mm} \urcorner$ means the proposition $\varphi$ will be placed into the pure Coq context, referred to in Figure 1 above the bold line. Notice in Figure \ref{assert_trans} that our subgoal denoted (1/2) has the pure goal $\ulcorner$x = z$\urcorner$. We use this tactic to turn this into a Coq goal so we can then use the hypotheses we currently have in our Coq context for our proof. Figure \ref{pure_intro} shows the state of our Coq context and subproof after the iPureIntro tactic. 
\begin{figure}[H]
\centering
  \includegraphics[width = 150mm]{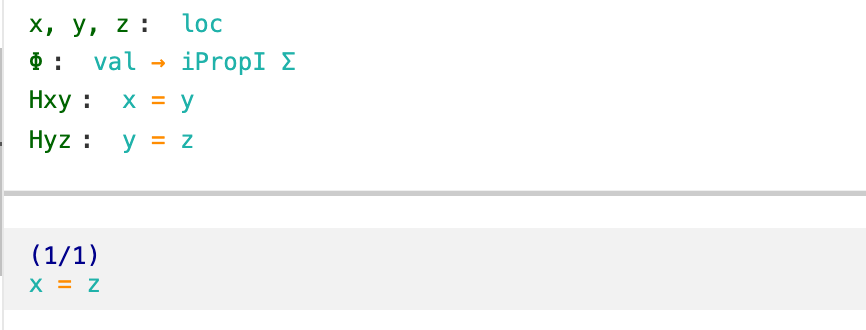}
  \caption{The result of using the tactic iPureIntro. $\ulcorner$x = z$\urcorner$ becomes a pure Coq goal so we can then use the hypotheses we currently have in our Coq context for our proof.}
  \label{pure_intro}
\end{figure}
    \item  $\textcolor{blue}{\textbf{iLeft}}$ : prove a disjunction $P \lor Q$ by proving the left side, P.
    \item $\textcolor{blue}{\textbf{iRight}}$ : prove a disjunction $P \lor Q$ by proving the right side, Q.
    \item  $\textcolor{blue}{\textbf{iSplitL "$H_1 ... H_n$"}}$ : splits a conjunction $P*Q$ into two proofs. The hypotheses $H_1 ... H_n$ are used for the left conjunct and the remaining hypotheses are moved to the right conjunct. Intuitionistic hypotheses are always available in both proofs. 
    \item  $\textcolor{blue}{\textbf{iSplitR "$H_0 ... H_n$"}}$ : splits a conjunction $P*Q$ into two proofs. The hypotheses $H_1 ... H_n$ are used for the right conjunct. This is the symmetric tactic to iSplitL. 
    \item $\textcolor{blue}{\textbf{iExists "$t_1, ..., t_n$"}}$ : provides a witness for an existential quantifier, $\exists x, ... t_1 ... t_n$.
\end{itemize}
\subsubsection{Elimination of logical connectives}
Some of the tactics in this category include:
\begin{itemize}
    \item $\textcolor{blue}{\textbf{iExFalso}}$ : changes the goal to proving False.
    \item $\textcolor{blue}{\textbf{iDestruct "H1" as ($x_1 ... x_n$) "H2"}}$ : eliminates a series of existential quantifiers in hypothesis H1 using Coq introduction patterns $x_1 ... x_n$ and names the resulting hypothesis H2. There are many special cases of iDestruct, here we introduce only the most basic usage. In the transitivity example presented above, we have the hypothesis "Heverything" that contains three separating conjunctions as seen in Figure \ref{desstruct1_intro}. The presence of these conjunctions tells us that we can further destruct this hypothesis, specifically into three individual hypotheses. We do this through the tactic iDestruct "Heverything" as "(Hxy \& Hyz \& Hxx)". This puts the three new hypotheses Hxy, Hyz, and Hxx into our proof state as seen in Figure \ref{destruct2_intro}.
\begin{figure}[H]
\centering
  \includegraphics[width = 150mm]{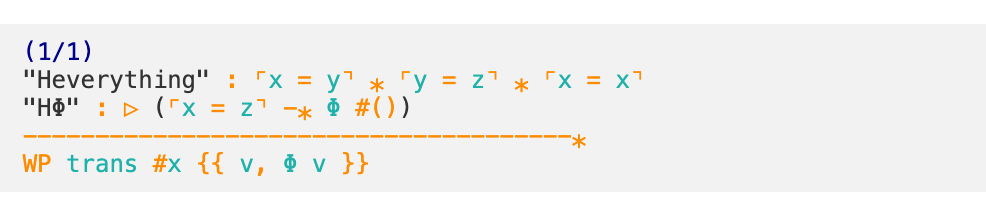}
  \caption{We have the hypothesis "Heverything" that contains three separating conjunctions}
  \label{desstruct1_intro}
\end{figure}
\begin{figure}[H]
\centering
  \includegraphics[width = 150mm]{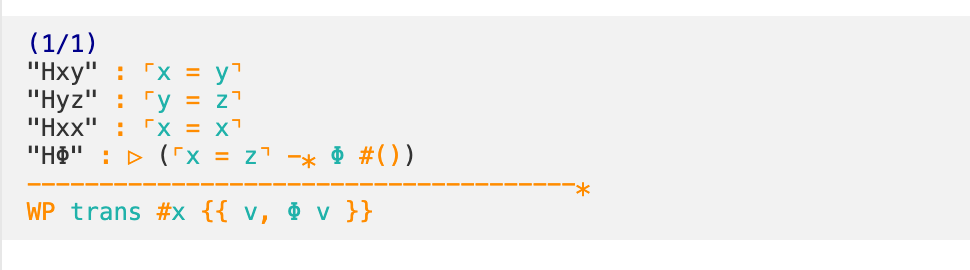}
  \caption{The result of using the tactic the tactic iDestruct "Heverything" as "(Hxy \& Hyz \& Hxx)". This puts the three new hypotheses Hxy, Hyz, and Hxx into our proof state.}
  \label{destruct2_intro}
\end{figure}
\end{itemize}
\subsubsection{Separation logic-specific tactics}
\begin{itemize}
    \item $\textcolor{blue}{\textbf{iFrame ($t_1 ... t_n$) "selpat"}}$ : cancels the Coq terms or Coq hypotheses $t_1 ... t_n$ and Iris hypotheses given by selpat in the goal.
    The constructs of the selection pattern have the following meaning:
    \begin{itemize}
        \item $\textcolor{blue}{\textbf{\%}}$ : repeatedly frame hypotheses from the Coq context.
        \item $\textcolor{blue}{\textbf{\#}}$ : repeatedly frame hypotheses from the intuitionistic context.
        \item $\textcolor{blue}{\textbf{*}}$ : frame as much of the spatial context as possible. Notice that framing spatial hypotheses makes them disappear, but framing Coq or intuitionistic hypotheses does not make them disappear. This tactic solves the goal if everything in the conclusion has been framed. 
    \end{itemize}
\end{itemize}
\subsubsection{Modalities}\label{iris-modals}
Some of the tactics in this category include:
\begin{itemize}
    \item $\textcolor{blue}{\textbf{iModIntro}}$ : introduces a modality in the goal. A deeper look into modalities occurred in Section \ref{modal}. In the transitivity example presented above, we will come along the notation "|=$\lbrace$ T $\rbrace$=>" which informs us that we have a modality as seen in Figure \ref{beforemod_intro}. We must introduce the modality into our goal, so we will use the tactic iModIntro as seen in Figure \ref{mod_intro}. This gets rid of our modality so we can continue with our proof. 
\begin{figure}[H]
\centering
  \includegraphics[width = 150mm]{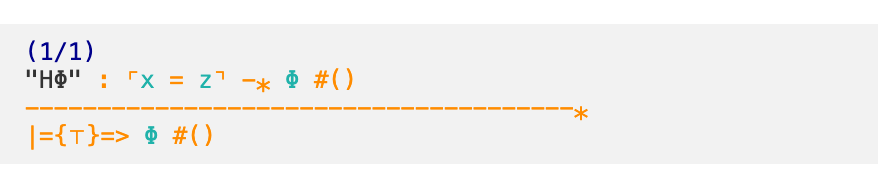}
  \caption{The notation "|=$\lbrace$ T $\rbrace$=>" informs us that we have a modality.}
  \label{beforemod_intro}
\end{figure}
\begin{figure}[H]
\centering
  \includegraphics[width = 150mm]{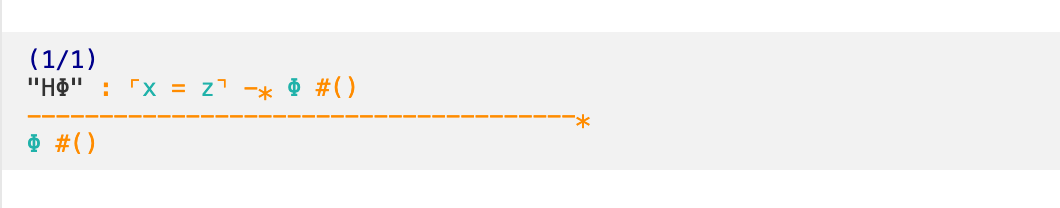}
  \caption{The result of using the tactic iModIntro. This tactic introduced the modality into our goal.}
  \label{mod_intro}
\end{figure}
    \item $\textcolor{blue}{\textbf{iNext}}$ : an alias of iModIntro ($\triangleright$ $^\wedge$\_ \_) which introduces the later modality. This eliminates a later in the goal and in exchange also strips one later from all the hypotheses.
\end{itemize}
\subsubsection{Iris}
\begin{itemize}
    \item $\textcolor{blue}{\textbf{iInv H as ($x_1 ... x_n$) "ipat"}}$ : opens an invariant in hypothesis H. The result is destructed using the Coq introduction patterns $x_1 ... x_n$ (for existential quantifiers) and then the proof mode introduction pattern ipat. 
\end{itemize}

\newpage
\section{Iris Proof Set-up}\label{set-up}
When you begin a proof in Iris, declarations need to be made in a mindful order to ensure that you are not overriding other modules imported later. In order to make sure the most relevant declarations and notations always take priority, import dependencies from furthest to closest. In particular, import modules in the following order: 
\begin{itemize}
    \item Coq
    \item stdpp
    \item iris.algebra
    \item iris.bi
    \item iris.proofmode
    \item iris.base\_logic
    \item iris.program\_logic
    \item iris.heap\_lang
\end{itemize}
We will see in Figure \ref{fig:dependencies}, for our example of a counter, that this order holds true for our dependencies. From these modules we export or import various files that will allow us to carry out our proof. For example, by requiring \framebox{From iris.proofmode Require Import tactics.}, we have access to files related to the interactive proof mode including the general tactics of the proof mode. For futher information, we refer readers to this \MYhref{https://gitlab.mpi-sws.org/iris/iris/-/blob/master/docs/proof_guide.md}{work-in-progress document} that explains further details about how Iris proofs are typically carried out in Coq.

\begin{figure}[H]
\centering
\begin{minipage}[t]{.80\linewidth}
\begin{lstlisting}[mathescape=true]
From iris.proofmode Require Import tactics.
From iris.base_logic.lib Require Export invariants.

From iris.program_logic Require Export weakestpre.
From iris.heap_lang Require Import proofmode.

From iris.heap_lang Require Import notation lang.
From iris.heap_lang.lib Require Import par.
\end{lstlisting}
\end{minipage}
\caption{Example of dependencies that are required for our proof that a counter is correct.}
 \label{fig:dependencies}
\end{figure}

\noindent Next, in order to complete our proof, we need to assume certain things about the instantiation of Iris. We section this portion of our file off by using the $\textcolor{blue}{\textbf{Section}}$ keyword. Once we are in this section, we need to specify that the Iris instantiation has sufficient structure to manipulate heaps and has ghost states available. We will require the following lines for our example proof: 
\begin{itemize}
    \item \framebox{Context `$\lbrace$!heapG $\Sigma \rbrace$.} : This line states that we assume the Iris instantiation has sufficient structure to manipulate the heap which means it will allow us to use the points-to predicate (discussed in Section \ref{sep-con}). 
    \item \framebox{Context `$\lbrace$ !spawnG $\Sigma \rbrace$.} : This line states that we have the necessary ghost states to prove the parallel composition construct defined in terms of fork which will be discussed in Section \ref{case-work}. 
    \item \framebox{Notation iProp := (iProp $\Sigma$).} : This line defines a shorthand notation for iProp $\Sigma$. The variable $\Sigma$ has to do with the ghost state available and the type of Iris propositions (Prop) that depend on this $\Sigma$. Since we will refer to the same $\Sigma$ throughout the whole file, we can define a shorthand notation that hides this $\Sigma$.
    \item \framebox{ Context (N : namespace).} : This line provides the namespace parameter for our invariant (discussed in Section \ref{invariants}) throughout the whole file. We will need an invariant to share access to a location. 
\end{itemize}

\noindent While we present these specifications for an Iris proof in terms of our specific example, everything will essentially be the same for most verifications. Therefore, it will help you create the start of your proof file no matter the specific problem at hand.

\newpage
\section{Case Study: Simple Counter}\label{beginning-example}
We will create one of the most simple program expressions to prove: \emph{incr}. \emph{incr} reads the value stored at heap location $n$ into a variable $(l= !n)$ and then writes $n+ 1$ back into the location. First we must define the program expression and program specs before we can prove they are correct. We determined these specifications in Section \ref{sep-formulas} and define them in Iris here. There are three different parts to defining and proving a program specification in Iris. We first have to define the program, denoted by the keyword $\textcolor{blue}{\textbf{Definition}}$. When defining your program, it may be useful to use the Iris Coq library which defines many notations for programming language constructs, such as lambdas, allocation, accessing and so on. The complete list of notations can be found in the \MYhref{https://gitlab.mpi-sws.org/iris/iris/-/blob/master/iris_heap_lang/notation.v}{iris-coq repository}. 

In this example the $\#$ notation denotes an embeded literal, variable, or number as a value of the programming language HeapLang. Our definition is followed by our $\textcolor{blue}{\textbf{Lemma}}$ and $\textcolor{blue}{\textbf{Proof}}$. The lemma we wish to prove here is that our program and the Hoare triple specifications is valid. We follow this lemma with the proof. In order to enter proof mode, we use the keyword $\textcolor{blue}{\textbf{Proof}}$ so that Coq knows we are ready to prove our lemma. 

\begin{figure}[H]
\centering
\begin{lstlisting}[mathescape=true]
Definition incr : expr := 
$\lambda$: "$l$", let: "n" := !"$l$" in
        "$l$" <- "n" + #1.

Lemma incr_spec ($\ell$ : loc) (n : Z):
{{{ $\ell$ $\mapsto$ #n }}} incr #$\ell$ {{{RET #(); $\ell$ $\mapsto$ #(n+1)}}}.

Proof.
iIntros ($\Phi$) "Hpt H$\Phi$".
wp_pures. wp_load. wp_let. wp_op.
wp_store. iModIntro.
iApply "H$\Phi$". iFrame.
Qed.
\end{lstlisting}
\end{figure}

\noindent In order to start proving the specifications in Figure \ref{fig:incr_start}, we first must unfold the triple notation, $\hoareTriple{x \mapsto n}{incr}{x \mapsto n + 1}$ , using the tactic $\textcolor{blue}{\textbf{iIntros ($\Phi$) "Hpt H$\Phi$"}}$. We see in Figure \ref{intros_incr} that this tactic introduces two hypotheses into the global context, changing our goal in the process. 

\begin{figure}[H]
  \includegraphics[width = 150mm]{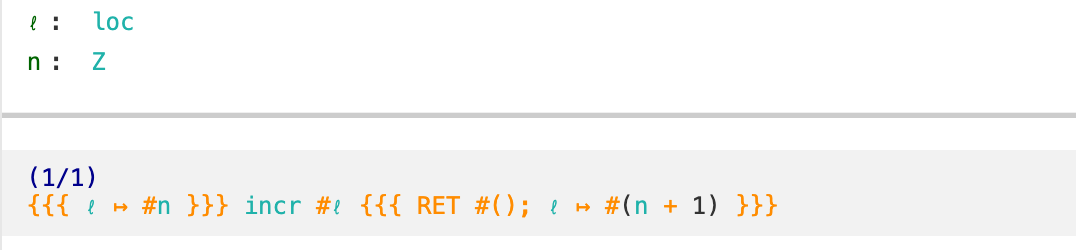}
  \centering
  \caption{The specifications of \emph{incr} that we must prove.}
  \label{fig:incr_start}
\end{figure}

\begin{figure}[H]
  \includegraphics[width = 150mm]{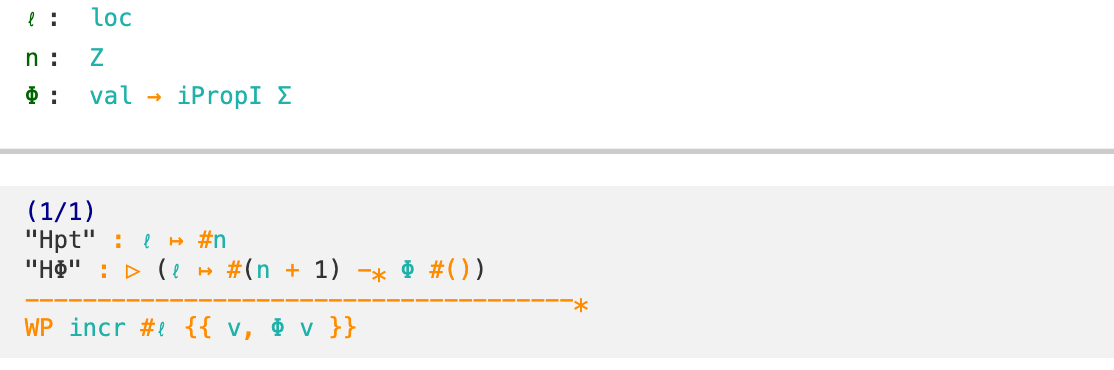}
  \centering
  \caption{The tactic iIntros ($\Phi$) "Hpt H$\Phi$" introduces two hypotheses into the spatial context; one from the precondition and one from the postcondition of the Hoare triple.}
  \label{intros_incr}
\end{figure}

\noindent In our new goal, we now see the "WP" notation which stands for weakest precondition. We know that HeapLang provides us with a lot of tactics that facilitate stepping through a HeapLang program as part of proving a weakest precondition. Therefore, we can look to Section \ref{heaplang} to see which tactics we could use. We use the tactic $\textcolor{blue}{\textbf{ wp\_pures}}$ which will perform as many pure reduction steps as possible as seen in Figure \ref{pures_incr}. 

\begin{figure}[H]
  \includegraphics[width = 150mm]{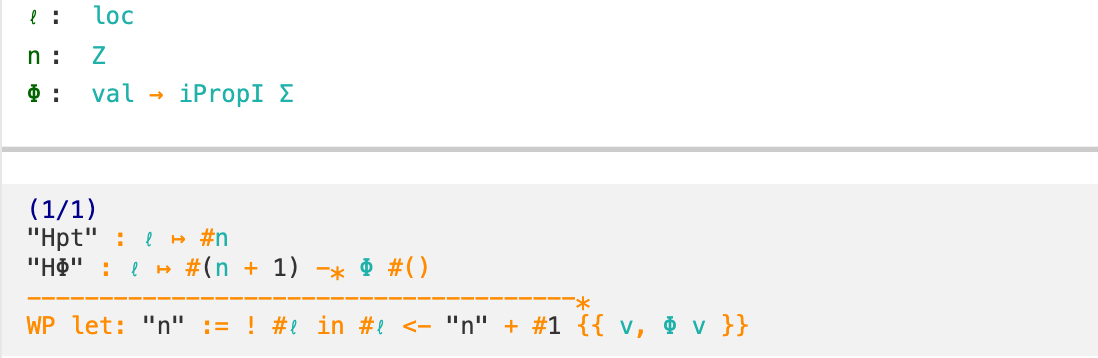}
  \centering
  \caption{The tactic wp\_pures performs as many pure reduction steps as possible.}
  \label{pures_incr}
\end{figure}

\noindent We now see not only a "WP" in our goal, but we also see a "let". Before we can deal with this "let" we must reduce the load operation. We use the tactic $\textcolor{blue}{\textbf{ wp\_load}}$ which automatically finds the points-to assertion in our spatial context. In this example, it finds the points-to assertion $\ell \mapsto \#n$. This means it will replace our $\ell$ with $n$ which can be seen in Figure \ref{load_incr}. \newline \newline 

\begin{figure}[H]
  \includegraphics[width = 150mm]{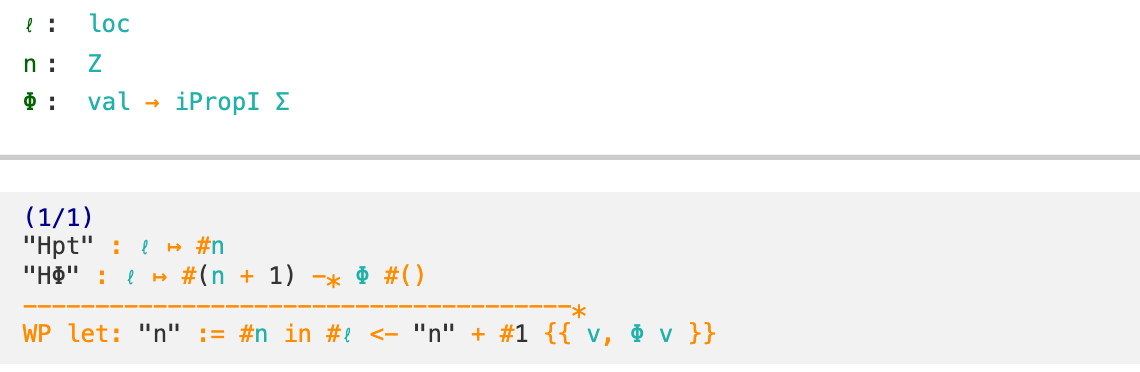}
  \centering
  \caption{The tactic wp\_load finds the points-to assertion $\ell \mapsto \#n$ and replaces $\ell$ with $n$.}
  \label{load_incr}
\end{figure}

\noindent We can now do some reductions. First, we reduce our let-binding through the tactic $\textcolor{blue}{\textbf{ wp\_let}}$ as seen in Figure \ref{let_incr}. 

\begin{figure}[H]
  \includegraphics[width = 150mm]{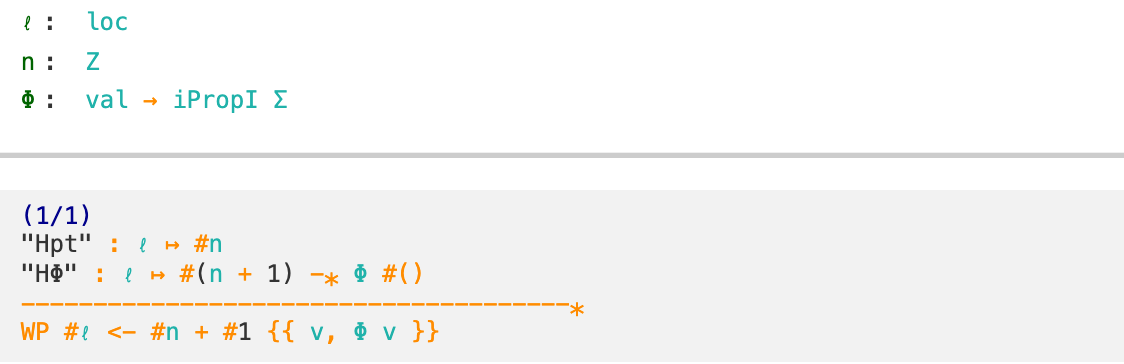}
  \centering
  \caption{The tactic wp\_let reduces the let-binding.}
  \label{let_incr}
\end{figure}

\noindent Then, we want to reduce $\#n + \#1$ to $\#(n+1)$ which can be done through the tactic $\textcolor{blue}{\textbf{ wp\_op}}$ which reduces unary or binary arithmetic operators as can be seen in Figure \ref{op_incr}.  

\begin{figure}[H]
  \includegraphics[width = 150mm]{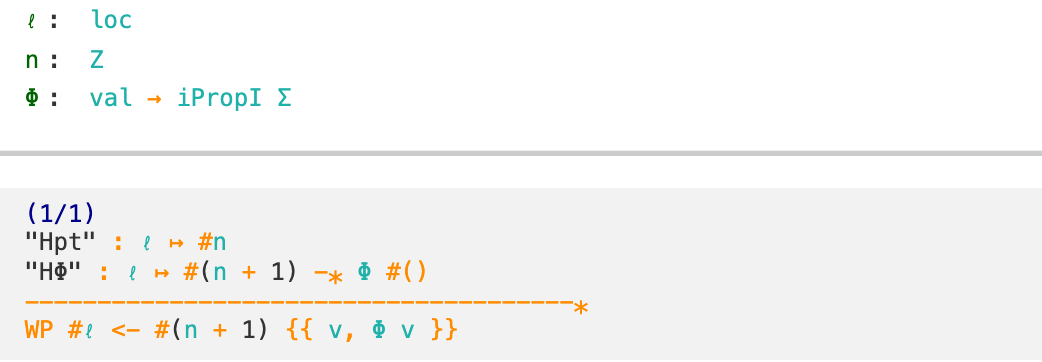}
  \centering
  \caption{The tactic wp\_op reduces $\#n + \#1$ to $\#(n+1)$.}
  \label{op_incr}
\end{figure}

\noindent We will now want to reduce our store operation. We use the tactic $\textcolor{blue}{\textbf{ wp\_store}}$ which automatically finds the points-to assertion in our spatial context. Once again the points-to assertion it finds is $\ell \mapsto \#n$ which can now be replaced with $\ell \mapsto \#(n+1)$ as seen in Figure \ref{store_incr}. 

\begin{figure}[H]
  \includegraphics[width = 150mm]{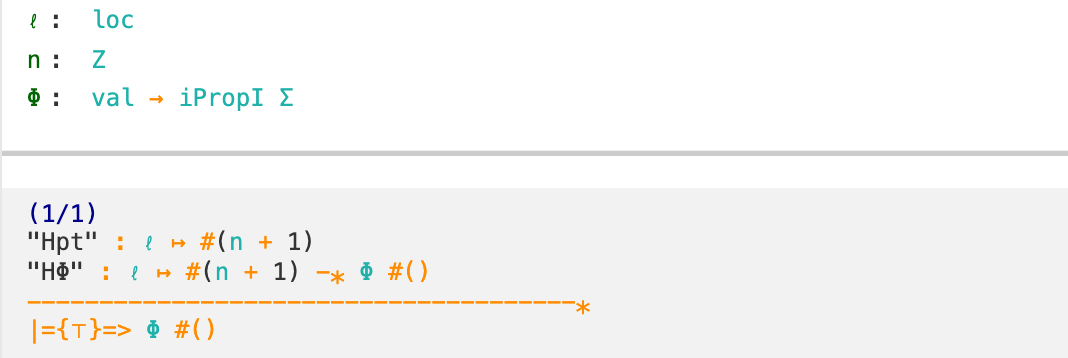}
  \centering
  \caption{The tactic wp\_store reduces the store operation. Find the points-to assertion $\ell \mapsto \#n$ and replaces it with $\ell \mapsto \#(n+1)$.}
  \label{store_incr}
\end{figure}

\noindent After we apply this tactic, we see the notation "|=$\lbrace$ T $\rbrace$=>" which informs us that we have a modality. We can refer to Section \ref{iris-modals} to determine which tactic we need to use next. Here we must introduce the modality into our goal, so we will use the tactic $\textcolor{blue}{\textbf{iModIntro}}$ as seen in Figure \ref{mod_incr}. 

\begin{figure}[H]
  \includegraphics[width = 150mm]{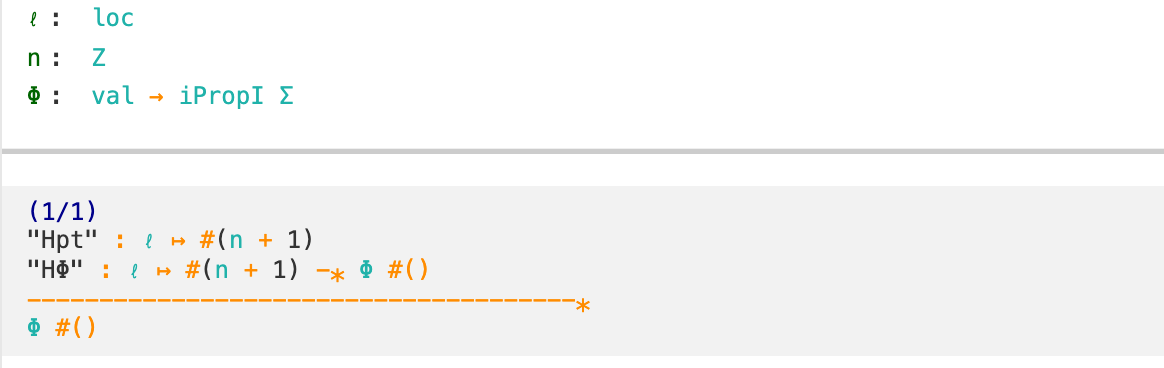}
  \centering
  \caption{The tactic iModIntro introduces the modality into our goal.}
  \label{mod_incr}
\end{figure}

\noindent We are left with "$\Phi \hspace*{2mm} \#()$" in our goal. If we look at our hypothesis "$H\Phi$" we see that we have both sides of the separating conjunction. The left side matches the hypothesis "Hpt" and the right hand side is our current goal. We can apply hypothesis "$H\Phi$" to our goal in order to reduce "$H\Phi$" to just the left side, $\ell \mapsto \#(n+1)$. We do this through the tactic $\textcolor{blue}{\textbf{iApply "$H\Phi$"}}$. This tactic matches the conclusion of the current goal against the conclusion of $H\Phi$ and generates goals for the premises of $H\Phi$. 

\begin{figure}[H]
  \includegraphics[width = 150mm]{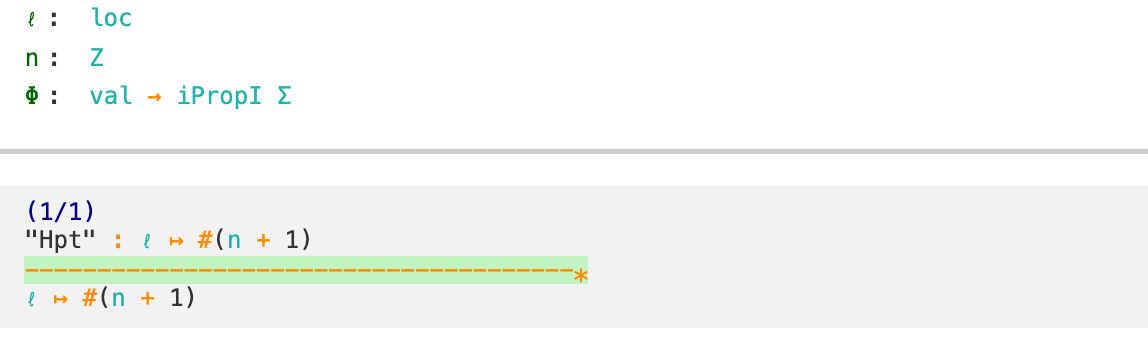}
  \centering
  \caption{The tactic iApply "$H\Phi$" matches the conclusion of the current goal against the conclusion of $H\Phi$ and generates goals for the premises of $H\Phi$. Our new proof is $\ell \mapsto \#(n+1)$.}
  \label{apply_incr}
\end{figure}

\noindent We see in Figure \ref{apply_incr} that this means our new proof will be  $\ell \mapsto \#(n+1)$. This new goal is the same as hypothesis "Hpt" which means we can easily finish this proof by cancelling out this hypothesis. The tactic $\textcolor{blue}{\textbf{iFrame}}$ allows us to cancel Coq hypothesis "Hpt" and finish our proof. We see the phrase $\textcolor{blue}{\textbf{"No more subgoals."}}$ which tells us we have completed the proof and can end it with the tactic $\textcolor{blue}{\textbf{Qed}}$. As discussed in Section \ref{proof-state}, $\textcolor{blue}{\textbf{Qed}}$ is how we inform Coq that our proof is done and exit proof mode.

\newpage
\section{Case Study: Parallel Counter}\label{case-work}
We will explain parts of this parallel increment example, however, to find the full code and explanation, refer to the \MYhref{https://gitlab.mpi-sws.org/iris/examples/-/blob/master/theories/lecture_notes/coq_intro_example_1.v}{Iris Project}. Before we dive into the example, let us provide some brief background knowledge on parallel threads. In the program we are going to define, we will use the commonly used parallel composition syntax $(e_1 || e_2)$ which denotes forking two threads that run $e_1$ and $e_2$ respectively and waiting until they complete. Parallel composition can be defined in terms of $\textbf{fork}$ and \lstinline`CAS`. \footnote{cite book} We will not go into the details of either term in this guide as it is not our main focus.  We now define our terms and the invariant we are going to use as the following: 

\begin{figure}[H]
\begin{lstlisting}[mathescape=true]
Definition counter ($\ell$ : loc) : expr := #$\ell$ <- !#$\ell$ + #1.

Definition counter_inv ($\ell$ : loc) (n : Z) : 
iProp := ($\exists$ (m : Z), $\ulcorner$(n $\leq$ m)%Z$\urcorner$ $*$ $\ell$ $\mapsto$ #m)%I.

\end{lstlisting}
\end{figure}

\noindent The body of our invariant is defined as \emph{counter\_inv}. Note that we use $\textcolor{blue}{\textbf{"\%I"}}$ to tell Coq to parse the logical formula as an Iris assertion; this means it will interpret the connectives as Iris connectives. We also use $\ulcorner ... \urcorner$ which is used to embed Coq propositions as Iris assertions. In this example, we make "$\leq$" an Iris assertion. In simple terms, this embedding means that the embedded assertion either holds for all resources or for none. We now define the specifications of the parallel counter we wish to prove. 

\begin{figure}[H]
\begin{lstlisting}[mathescape=true]
Lemma parallel_counter_spec ($\ell$ : loc) (n : Z):
    {{{ $\ell$ $\mapsto$ #n }}} (counter $\ell$) ||| (counter $\ell$) ;; !#$\ell$ 
            {{{m, RET #m; $\ulcorner$(n $\leq$ m)%Z$\urcorner$ }}}.
\end{lstlisting}
\end{figure}

\noindent Here we will  describe the tactics that effect our invariant. We first want to allocate an invariant (in namespace $N$) and transfer the points-to-predicate into it. This can be achieved by using the tactic $\textcolor{blue}{\textbf{inv\_alloc}}$. The allocation of an invariant involves the fancy update modality which means we also must use the $\textcolor{blue}{\textbf{iMod}}$ tactic around the lemma. The $\textcolor{blue}{\textbf{iMod}}$ tactic knows about the structural rules of the update modalities and the interaction of the update modality and the weakest precondition assertion and therefore it will automatically remove the modality as much as possible. Our full tactic for this purpose becomes $\textcolor{blue}{\textbf{iMod (inv\_alloc N \_ (incr\_inv $\ell$ n) with "[Hpt]") as "\#HInv"}}$. The tactic $\textcolor{blue}{\textbf{inv\_alloc}}$ takes three parameters, the namespace in which to allocate, the mask used on the fancy update modality (not important for our purposes so here we leave it implicit by using \_ ), and the body of the invariant to be allocated. We see this tactic also contains the "with" construct. This tells $\textcolor{blue}{\textbf{iMod}}$ which of the assumptions are to be used to satisfy the assumptions of the $\textcolor{blue}{\textbf{inv\_alloc}}$ rule. Finally, the last part of this tactic is "as "\#Hinv"". This tells the $\textcolor{blue}{\textbf{iMod}}$ tactic to name the conclusion of the $\textcolor{blue}{\textbf{inv\_alloc}}$ rule "Hinv" and add it to the persistent hypotheses; the \# notation is used to denote persistent hypotheses. 
\newline \newline
\noindent In general terms of the structure of this proof, we would have three subgoals. We would have two proofs to show each thread does the correct thing, and a third goal to show that the combined conclusion of the two threads implies the desired conclusion. Let us first consider the proof that the first thread is correct. The expression is not atomic, so we cannot open the invariant immediately. Thus we must first use the bind rule to reshape it, then we can open the invariant. We can see the state of our proof in Figure \ref{inv1} right before we open our invariant. The next portion of this proof will show how to open and close this invariant.

\begin{figure}[H]
  \includegraphics[width = 150mm]{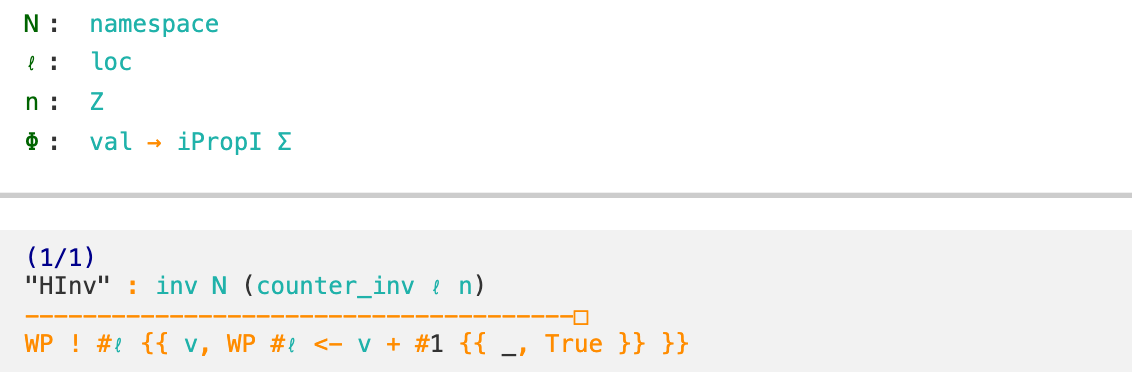}
  \centering
  \caption{The proof state before we open the invariant. The last tactic used before this process was wp\_bind (!\#$\ell$)\%E.}
  \label{inv1}
\end{figure}

\noindent We can open the invariant to read a value stored at location $\ell$. We open an invariant by using tactic $\textcolor{blue}{\textbf{iInv}}$. This tactic takes a namespace as the first argument and two named assertions. There are two parts of the $\textcolor{blue}{\textbf{inv\_open}}$ rule because there are two parts to the conclusion. Therefore, we use the full tactic $\textcolor{blue}{\textbf{iInv N as "H" "Hclose"}}$ as can be seen in Figure \ref{inv2}. 

\begin{figure}[H]
  \includegraphics[width = 150mm]{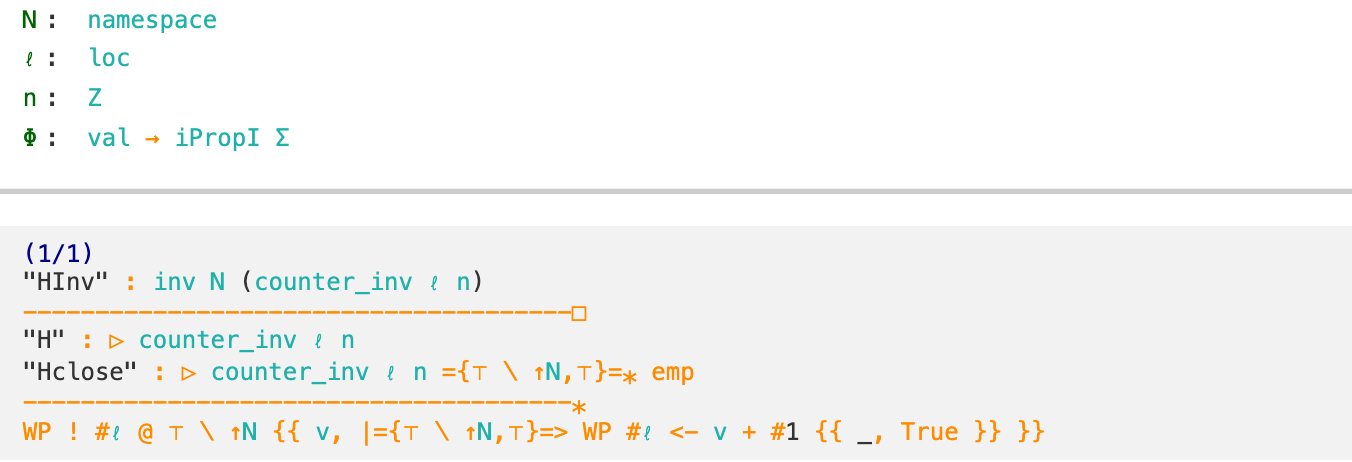}
  \centering
  \caption{We use tactic iInv N as "H" "Hclose" to open the invariant so that we have access to the resource and can read the value of it.}
  \label{inv2}
\end{figure}

\noindent We now can read the value since we have the resources. However, if we look at the \emph{incr\_inv} assertion, we see we need to eliminate our existential "H" to get a points-to predicate before we can read the memory location. We can do this through the $\textcolor{blue}{\textbf{iDestruct}}$ tactic which will eliminate the existential quantifier in the hypothesis. We will not go into details here as to the rest of the syntax for this tactic, but this tactic will provide us with a points-to assertion in context as seen in Figure \ref{inv3}.

\begin{figure}[H]
  \includegraphics[width = 150mm]{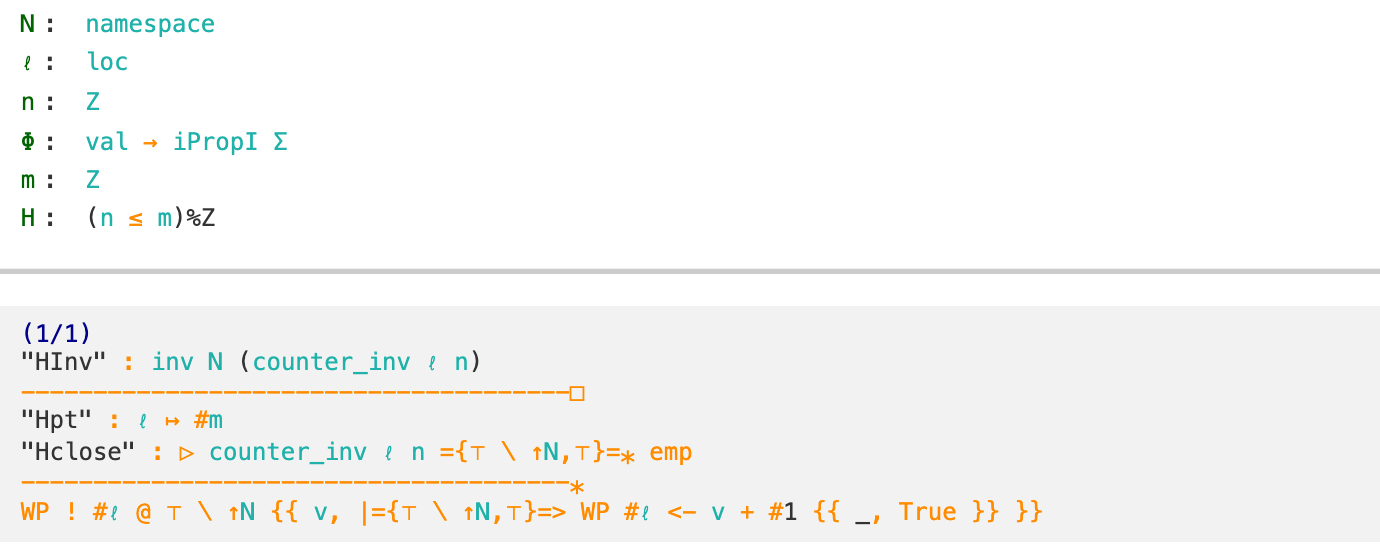}
  \centering
  \caption{We use tactic iDestruct "H" as (m) ">[\% Hpt]" to eliminate the existential quantifier in the hypothesis.}
  \label{inv3}
\end{figure}

\noindent We can then use the $\textcolor{blue}{\textbf{wp\_load}}$ tactic to read the memory location in Figure \ref{inv4}.

\begin{figure}[H]
  \includegraphics[width = 150mm]{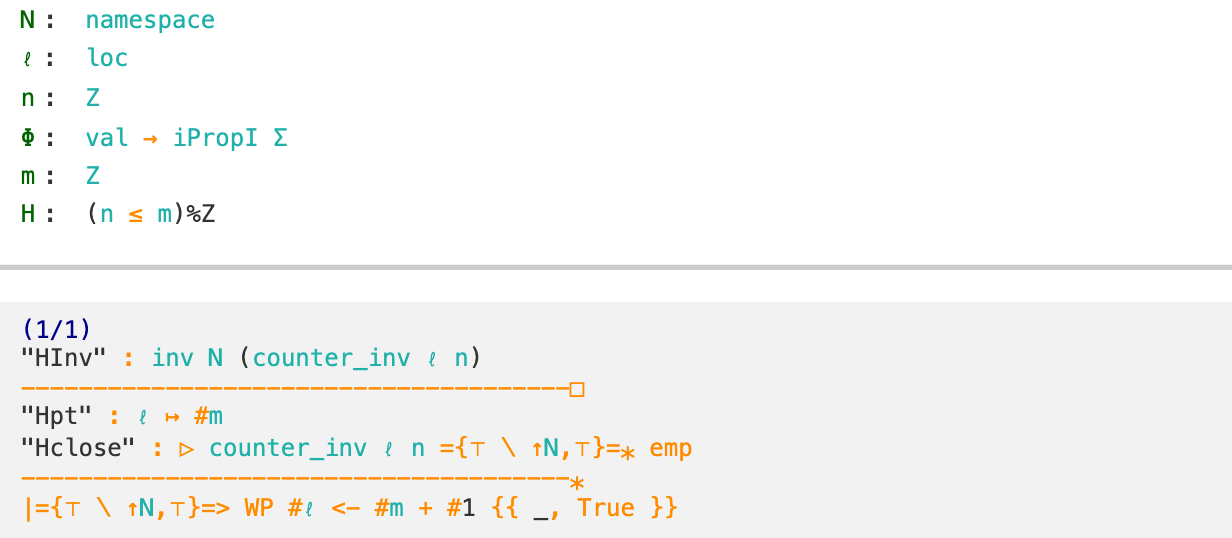}
  \centering
  \caption{We use tactic wp\_load to read the memory location.}
  \label{inv4}
\end{figure}

\noindent Now that we have read our value, we must close the invariant. We use the "Hclose" assertion that we got when we opened the invariant to now close it. We close the invariant by transferring the points-to predicate back. Closing the invariant involves manipulation of fancy update modalities so we must use the $\textcolor{blue}{\textbf{iMod}}$ tactic. We do not care about the conclusion of the "Hclose" assertion in this case, so we will use the "\_" notation to state we can ignore it. Our full tactic will be $\textcolor{blue}{\textbf{iMod("Hclose" with "[Hpt]") as "\_"}}$ whose results can be seen in Figure \ref{inv5}.

\begin{figure}[H]
  \includegraphics[width = 150mm]{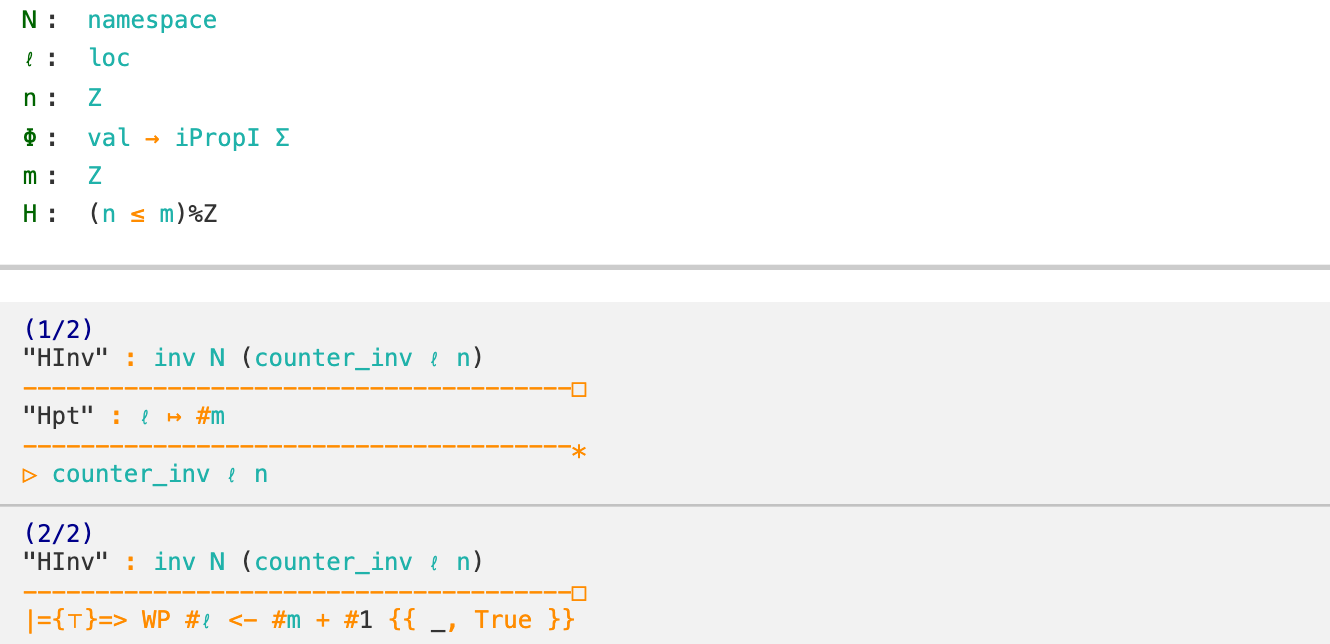}
  \centering
  \caption{We use tactic iMod("Hclose" with "[Hpt]") as "\_" to close the invariant.}
  \label{inv5}
\end{figure}

\noindent As can be seen in Figure \ref{inv5} above, we now have a subgoal we must prove before we are done with closing our invariant. Our subgoal, denoted by (1/2), starts with the later modality $\triangleright$ which tells us we need to use the tactic $\textcolor{blue}{\textbf{iNext}}$. This tactic will introduce the later modality, eliminating it from our goal as seen in Figure \ref{next-parallel}. 

\begin{figure}[H]
  \includegraphics[width = 150mm]{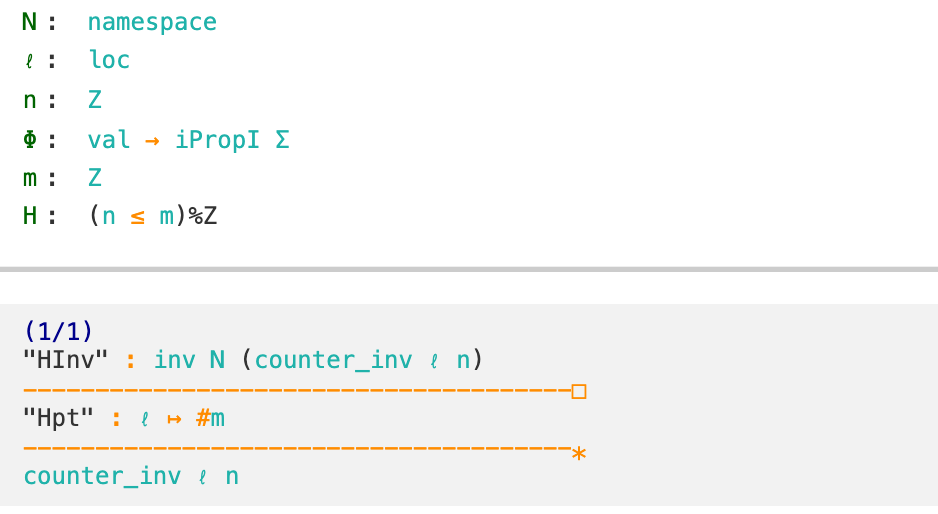}
  \centering
  \caption{We use tactic: iNext to introduce the later modality and eliminate it from our goal.}
  \label{next-parallel}
\end{figure}

\noindent We must also introduce $m$ into our goal which is possible through the tactic $\textcolor{blue}{\textbf{iExists $m$}}$ since there is an existential quantifier in our invariant. The tactic $\textcolor{blue}{\textbf{iExists}}$ provides a witness for an existential quantifier, in this case ($\exists$ (m : Z)) as seen in Figure \ref{exists-parallel}. 

\begin{figure}[H]
  \includegraphics[width = 150mm]{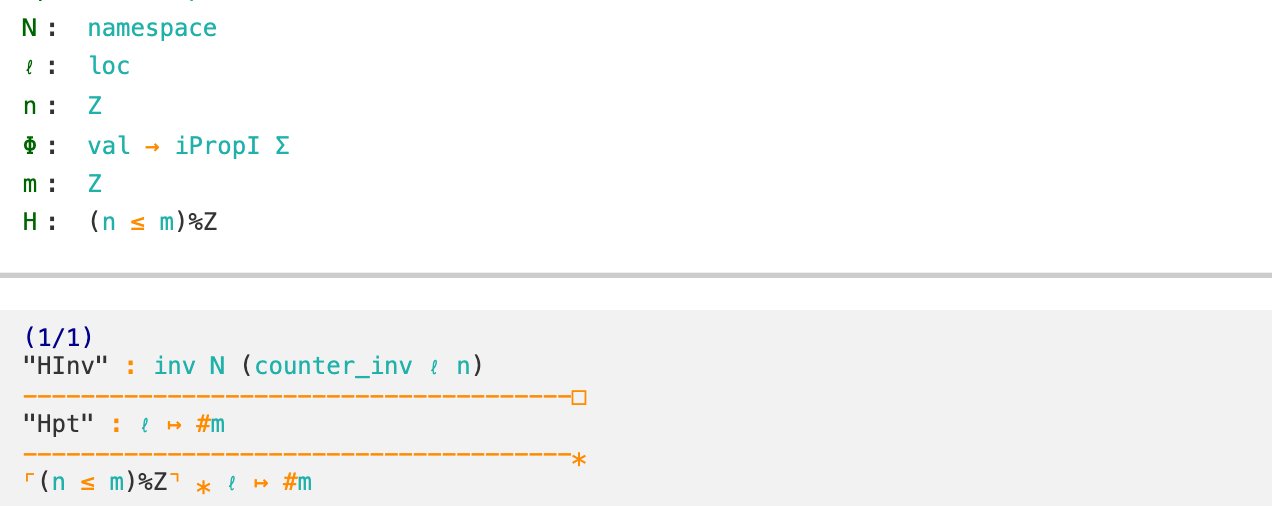}
  \centering
  \caption{We use tactic: iExists m to provide a witness for the existential quantifier ($\exists$ (m : Z)).}
  \label{exists-parallel}
\end{figure}

\noindent At this point we see we have half of the separating conjunction, meaning we can use the tactic $\textcolor{blue}{\textbf{iFrame}}$ to cancel the right side of the Coq term in the goal as done in Figure \ref{frame-parallel}.

\begin{figure}[H]
  \includegraphics[width = 150mm]{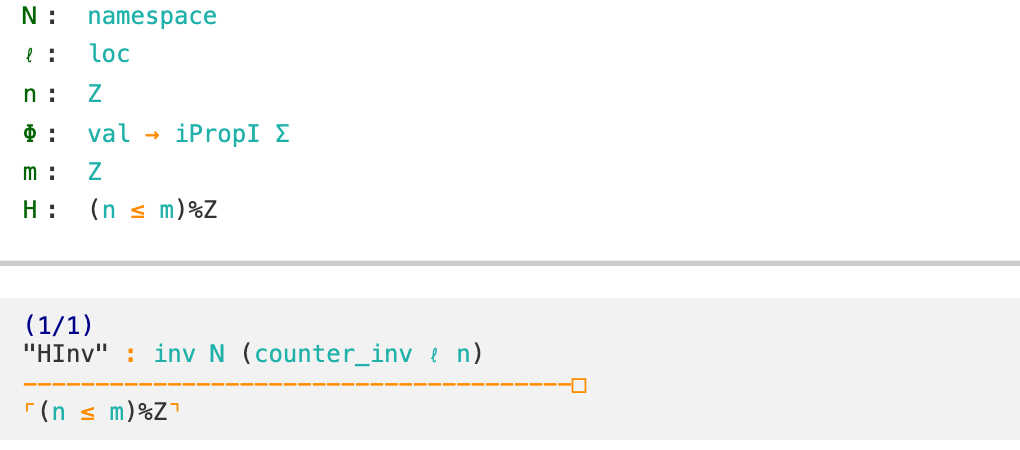}
  \centering
  \caption{We use tactic: iFrame to cancel the right side of the Coq term in the goal.}
  \label{frame-parallel}
\end{figure}

\noindent We can see we are almost finished with this subproof, however our goal is wrapped in $\ulcorner ... \urcorner$ which means it is in pure Coq context and we must introduce a pure goal. We do this through the notation "!\%" and the tactic $\textcolor{blue}{\textbf{iIntros}}$, giving us the full tactic $\textcolor{blue}{\textbf{iIntros "!\%"}}$. We then see our goal in Figure \ref{intros-parallel} perfectly matches the Coq hypothesis "H" in our global context. 

\begin{figure}[H]
  \includegraphics[width = 150mm]{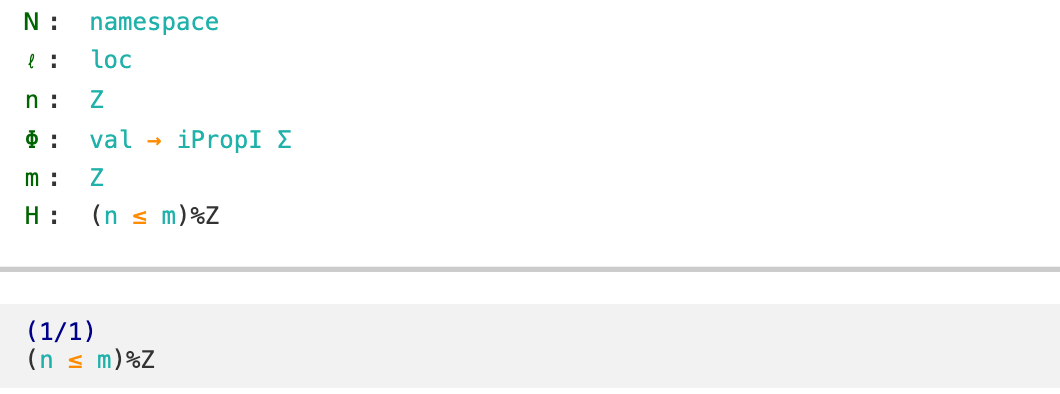}
  \centering
  \caption{We use tactic iIntros "!\%" to introduce $\ulcorner (n \leq m)\%Z \urcorner$ as a pure Coq goal. Note that the tactic iPureIntro would also work here.}
  \label{intros-parallel}
\end{figure}

\noindent We are able to use the tactic $\textcolor{blue}{\textbf{auto}}$ which will automatically find this connection and finish our subproof for us. We know this is the case when we see the phrase $\textcolor{blue}{\textbf{"There are unfocused goals."}}$. This means we can continue our main proof. All of these tactics were described one after each other because they could be linked into one big tactic. We can use ";" to string together tactics as follows:$\textcolor{blue}{\textbf{iNext; iExists m; iFrame; iIntros "!\%"; auto.}}$ This one tactic will have the same effect as all 4 tactics being individually called.

\noindent This fully concludes the opening and closing of the invariant which may have to be done multiple times throughout our proof. In total, this process required the chunk of code below. While this was specific to our parallel increment example, the tactics used to open invariants in most cases will look similar. 

\begin{figure}[H]
   \begin{lstlisting}
    iInv N as "H" "Hclose".
    iDestruct "H" as (m) ">[% Hpt]".
    wp_load.
    iMod ("Hclose" with "[Hpt]") as "_".
      { iNext. iExists m. iFrame. iIntros "!%". auto. }
   \end{lstlisting}
\end{figure}

\newpage
\section{Case Study: Bank}
We will explain parts of a bank example, however, we want to note that this is a simplified version of the bank implemented in: \MYhref{https://plv.csail.mit.edu/blog/iris-intro.html}{A brief introduction to Iris}. Our version does not include locks in our bank for simplicity. But, we provide a deeper explanation into the Iris tactics being used, especially in terms of ghost states. For interested readers, we highly suggest working through the full, more advanced code in that guide, upon total understanding of this version of the bank. 
\newline \newline
\noindent Before we dive into the definitions for this example, we need to add different declarations than we presented in Section \ref{set-up}. Here we must take into account our invariants and ghost states, so we add the following dependencies to our proof:

\begin{figure}[H]
\begin{lstlisting}
From iris.algebra Require Import lib.excl_auth.
From iris.heap_lang Require Import proofmode notation.
From iris.base_logic.lib Require Export invariants.
Set Default Proof Using "All".
\end{lstlisting}
\end{figure}

\noindent We also need to specify the cameras/RAs used as ghost states in this proof. To do so we add the following line to go with our other contexts: 

\[
   \framebox{Context $`\lbrace$ inG $\Sigma$ (authR (optionUR (exclR ZO)))$\rbrace$ $`\lbrace$OfeDiscrete ZO$\rbrace$.}
\]

\noindent Now we can begin defining our program. We will need two main parts for this program: the bank with our accounts and the ability to transfer money from one account to another. Once we have these, we will also need to formulate our invariants. We will need one invariant to be associated with the account (account\_inv), it will tie the account balance to a ghost variable and take the location where the account balance is stored. The other invariant will be associated with the bank (bank\_inv), it will hold the fragments of the account balances and states that the account balances sum to zero.
\newline \newline
\noindent First, let us define our bank. Our function \emph{bank} will construct a bank with two accounts, both with a zero balance: 
\begin{figure}[H]
\begin{lstlisting}[mathescape=true]
 Definition bank: val :=
  $\lambda$: <>,
  let: "a_bal" := ref #0 in
  let: "b_bal" := ref #0 in
  ( "a_bal" , "b_bal").
\end{lstlisting}
\end{figure}

\noindent We now define our \emph{transfer} function. The function \emph{transfer} will move money from the first to the second account. Note that we do not check that there is enough money in the account before transferring the money, meaning negative balances are allowed. Here we verify only that this function is safe.

\begin{figure}[H]
\begin{lstlisting}[mathescape=true]
Definition transfer: val :=
 $\lambda$: "bank" "amt",
 let: "a" := Fst "bank" in
 let: "b" := Snd "bank" in
 Snd "a" <- !(Snd "a") - "amt";;
 Snd "b" <- !(Snd "b") + "amt";;
 #().
\end{lstlisting}
\end{figure}

\noindent Now that we have the basic setup of our bank, we can define our invariants as described above: 

\begin{figure}[H]
\begin{lstlisting}[mathescape=true]
Definition bank_inv ($\gamma_1$ $\gamma_2$: gname): iProp :=
  ($\exists$ (bal1 bal2: Z),
   own $\gamma_1$ ($\bullet$E bal1) $\text{*}$ 
   own $\gamma_2$ ($\bullet$E bal2) $\text{*}$ 
   $\ulcorner$(bal1 + bal2)%Z = 0$\urcorner$)%I.
   
Definition account_inv $\gamma$ bal_ref : iProp :=
  $\exists$ (bal: Z), bal_ref $\mapsto$ #bal $\text{*}$  own $\gamma$ ($\circ$E bal).
\end{lstlisting}
\end{figure}

\noindent In our invariant we introduce ghost variables. In order to reason about these later, we must also introduce the function \emph{ghost\_var\_alloc}. In these ghost states, \emph{a} represents an element of arbitrary type. The ghost state $\textcolor{blue}{\textbf{own $\gamma_1$ ($\bullet$E bal)}}$ represents "authoritative ownership" and the ghost state
$\textcolor{blue}{\textbf{own $\gamma$ ($\circ$E bal)}}$ represents "fragmentary ownership", however, both have exclusive ownership meaning they are symmetric. In general terms, the state that has authoritative ownership will go to the global invariant and the state that has fragmentary ownership will be handed out to other invariants (such as lock invariants). By using the function \emph{ghost\_var\_alloc} we can allocate the pair of ghost states. We will not  explain the function \emph{ghost\_var\_alloc} any further, but we will show how to use it.  

\begin{figure}[H]
\begin{lstlisting}[mathescape=true]
Lemma ghost_var_alloc `{inG $\Sigma$ (authR (optionUR (exclR A)))} 
    `{OfeDiscrete A} (a: A) :
          $\vdash$ |==> $\exists$ $\gamma$, own $\gamma$ ($\bullet$E a) $\text{*}$ own $\gamma$ ($\circ$E a).
Proof.
  iMod (own_alloc ($\bullet$E a $\cdot$ $\circ$E a)) as ($\gamma$) "[H1 H2]".
  { apply excl_auth_valid. }
  iModIntro. iExists $\gamma$. iFrame.
Qed.
\end{lstlisting}
\end{figure}

\noindent The last piece we need to define before we can prove our theorem is a way to tie together each account and the bank. We do this through \emph{is\_bank}:

\begin{figure}[H]
\begin{lstlisting}[mathescape=true]
Definition is_bank (b: val): iProp :=
  ($\exists$ (acct1 acct2: val) ($\gamma_1$ $\gamma_2$: gname),
  $\ulcorner$ b = (acct1, acct2)%V$\urcorner$ $\text{*}$  
  is_account acct1 $\text{*}$ 
  is_account acct2 $\text{*}$ 
  inv N (bank_inv $\gamma_1$ $\gamma_2$)).
\end{lstlisting}
\end{figure}

\noindent The proof we will go through in detail for this example is the function, \emph{bank}. The function \emph{bank} has to create all of the ghost states, invariants, and argue all of these initially hold. Therefore, we will explain this code in great detail to give some insight on how to use ghost states and the tactics related to them.
\newline \newline 
\noindent We need to define our program specifications for the theorem we are going to prove, \emph{wp\_bank}. Similar to our other examples, we can do this through a Hoare triple. Our Hoare triple for this program will be $\hoareTriple{True}{bank}{\exists b, RET \hspace*{2mm} b;\hspace*{1mm} is\_bank \hspace*{2mm} b}$. Note that here, \emph{RET b} means that we return $b$ in the postcondition. We can write this in Iris logic as follows: 
\begin{figure}[H]
\begin{lstlisting}[mathescape=true]
 Theorem bank_spec :
  {{{ True }}}
    bank #()
  {{{ b, RET b; is_bank b }}}.
\end{lstlisting}
\end{figure}

\noindent We start off our proof with two simple tactics. We first unfold our Hoare Triple through the tactic $\textcolor{blue}{\textbf{iIntros ($\Phi$) "\_ H$\Phi$"}}$ and then we want to unfold the function \emph{bank} through the tactic $\textcolor{blue}{\textbf{wp\_rec}}$. This leaves us with the proof state seen below in Figure \ref{intros-bank}. 

\begin{figure}[H]
  \includegraphics[width = 140mm]{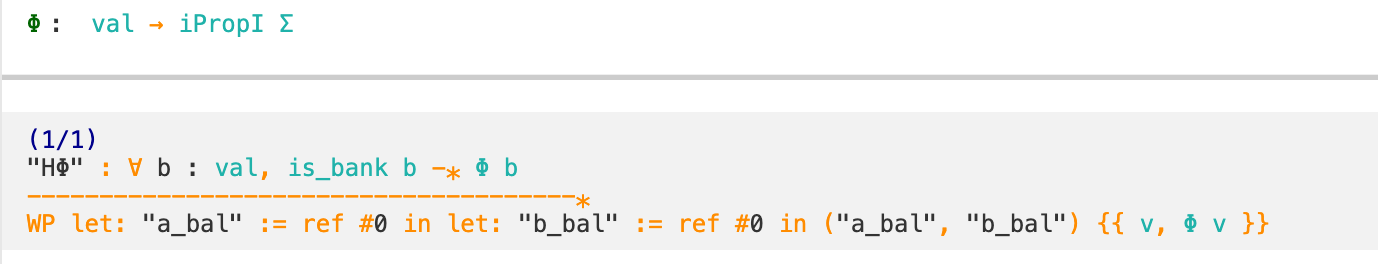}
  \centering
  \caption{Our proof state after unfolding our Hoare triple and the function \emph{bank}.}
  \label{intros-bank}
\end{figure}

\noindent In our goal we see we have two references, "a\_bal" := ref \#0 and "b\_bal" := ref \#0, one for each account. We want to pull these out of our goal and into our spatial context so we can use them. In order to do this, we can use the HeapLang tactic $\textcolor{blue}{\textbf{wp\_alloc l as "H"}}$. This tactic reduces the allocation (reference) instruction and allows us to put it at a new location. Our full tactic here will be $\textcolor{blue}{\textbf{wp\_alloc a\_ref as "Ha"}}$. This tells Iris we are taking the reference of "a\_bal" (since it occurs first in the goal) and placing it at location "a\_ref" in our Coq context and naming its points-to assertion "Ha" in our spatial context. Similarily, we will use tactic $\textcolor{blue}{\textbf{wp\_alloc b\_ref as "Hb"}}$ to do the same for our second reference "b\_bal". This brings us to the proof state seen in Figure \ref{alloc-bank}.

\begin{figure}[H]
  \includegraphics[width = 150mm]{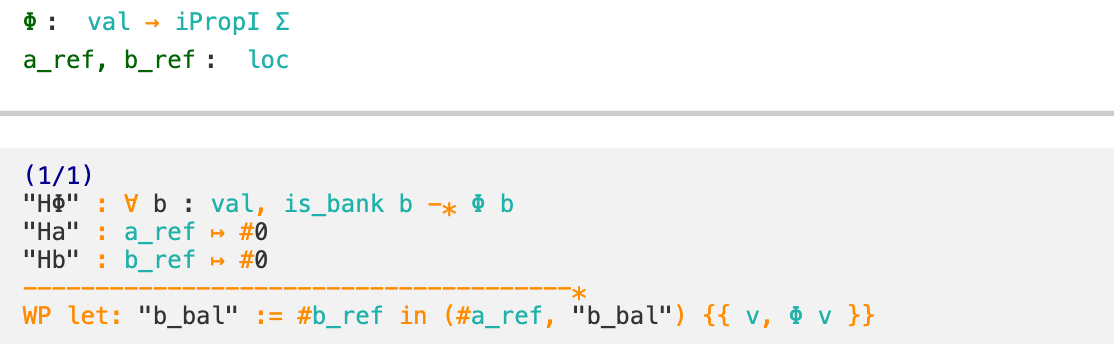}
  \centering
  \caption{Our proof state after reducing our references for each account balance through the tactics wp\_alloc a\_ref as "Ha" and wp\_alloc b\_ref as "Hb".}
  \label{alloc-bank}
\end{figure}

\noindent We are now at the point of the proof where we need to introduce our ghost variables. We will execute the ghost variable change using the function \emph{ghost\_var\_alloc} and destruct it as ghost variable $\gamma_1$ with names "Hown1" and "H$\gamma_1$". Here "Hown1" and "H$\gamma_1$" name the two resources $\textcolor{blue}{\textbf{own $\gamma$ ($\bullet$E a)}}$ and $\textcolor{blue}{\textbf{own $\gamma$ ($\circ$E a)}}$ respectively. As mentioned earlier, \emph{a} represents an element of arbitrary type, in our specific example $a$ will be replaced with $0$. In our function, we can allocate a new ghost variable, under an update modality because it requires modifying the global ghost state as seen below \cite{mit}. 
\newline \newline

\noindent As we already discussed above, we need to execute the ghost variable change. To do so, we use the tactic $\textcolor{blue}{\textbf{iMod (ghost\_var\_alloc (0: ZO)) as ($\gamma_1$) "(Hown1 \& H$\gamma_1$)"}}$ which introduces $\textcolor{blue}{\textbf{own $\gamma_1$ ($\bullet$E 0)}}$ and $\textcolor{blue}{\textbf{own $\gamma_1$ ($\circ$E 0)}}$ as seen in Figure \ref{ghost1}. Similarly, we do the exact same thing for parallel results for our second ghost variable $\gamma_2$.

\begin{figure}[H]
  \includegraphics[width = 140mm]{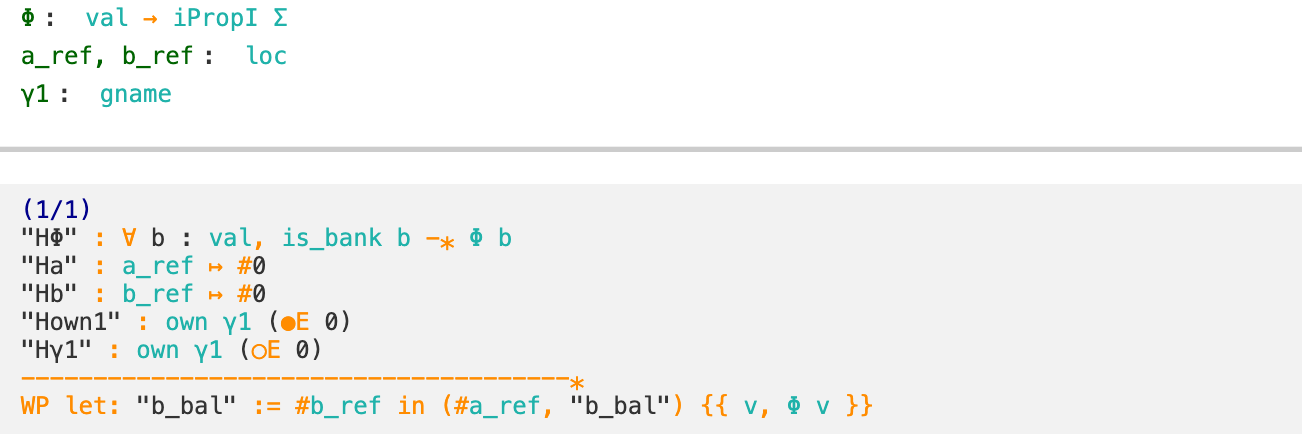}
  \centering
  \caption{Proof state after using tactic iMod (ghost\_var\_alloc (0: ZO)) as ($\gamma_1$) "(Hown1 \& H$\gamma_1$)". Similarly, we do the exact same thing for parallel results for our second ghost variable $\gamma_2$. }
  \label{ghost1}
\end{figure}

\noindent Now that we have introduced our ghost states, we can allocate the \emph{bank\_inv} invariant. Recall that in \emph{bank\_inv} we say the balances of the two accounts sums to zero. We will show that this statement initially holds true. In order to allocate our invariant we must use the tactic $\textcolor{blue}{\textbf{inv\_alloc}}$. Remember from the parallel increment example that $\textcolor{blue}{\textbf{inv\_alloc}}$ takes three parameters, the namespace in which to allocate, the mask used on the fancy update modality (not important for our purposes so here we leave it implicit by using \_ ), and the body of the invariant to be allocated. As stated before, the allocation of an invariant also involves the fancy update modality which means we must use the $\textcolor{blue}{\textbf{iMod}}$ tactic around the lemma. Therefore, our full tactic results $\textcolor{blue}{\textbf{iMod (inv\_alloc N \_ (bank\_inv $\gamma_1 \gamma_2$) with "[Hown1 Hown2]") as "Hinv"}}$. The resulting proof state seen in Figure \ref{inv-bank} requires us to finish off the subgoal this creates. This subproof will be only five tactics, all of which we have explained in detail elsewhere in this guide. We suggest attempting this small proof on your own, before referring to the code below. 

\begin{figure}[H]
  \includegraphics[width = 140mm]{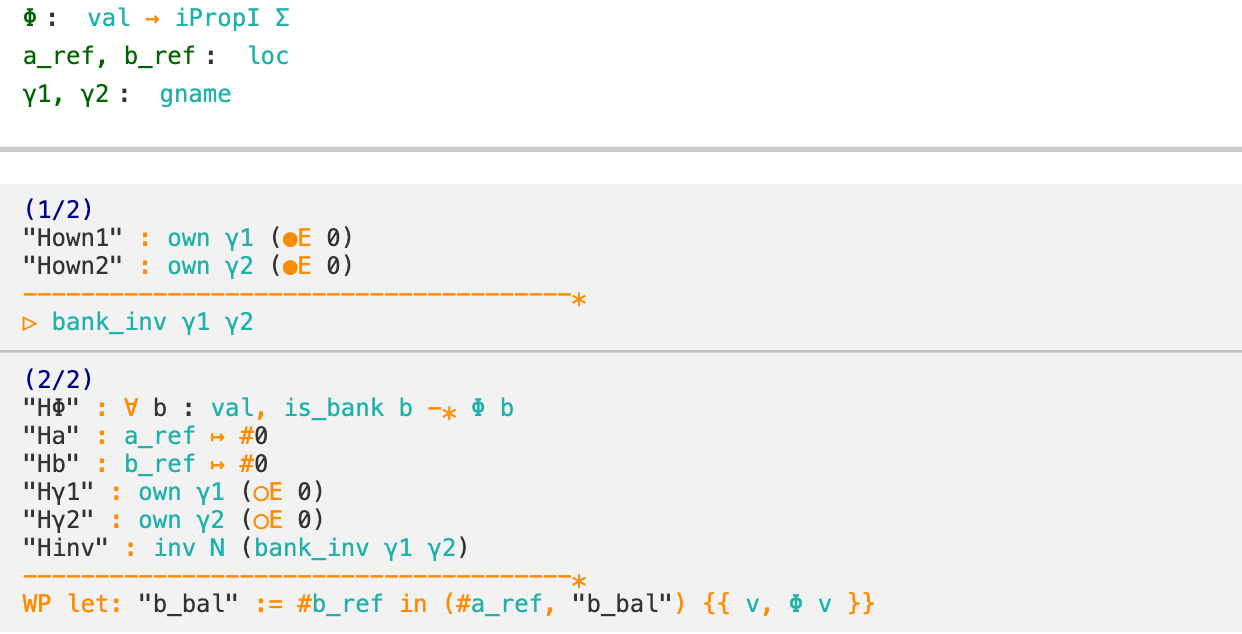}
  \centering
  \caption{Proof state after using tactic iMod (inv\_alloc N \_ (bank\_inv $\gamma_1 \gamma_2$) with "[Hown1 Hown2]") as "Hinv". }
  \label{inv-bank}
\end{figure}

\noindent The rest of the proof is straightforward, using many of the tactics we have already explained in great detail. The full code for the proof is posted below, however we will not go into detail about each tactic. While working out the rest of this example, we highly suggest referring to Section \ref{iris-discussion} to see all of the tactics available to you.

\begin{figure}[H]
\begin{lstlisting}[mathescape=true]
Proof.
iIntros ($\Phi$) "_ H$\Phi$".
wp_rec. 
wp_alloc a_ref as "Ha".
wp_alloc b_ref as "Hb".
iMod (ghost_var_alloc (0: ZO)) as ($\gamma_1$) "(Hown1 & H$\gamma_1$)".
iMod (ghost_var_alloc (0: ZO)) as ($\gamma_2$) "(Hown2 & H$\gamma_2$)".
iMod (inv_alloc N _ (bank_inv $\gamma_1$ $\gamma_2$)
with "[Hown1 Hown2]") as "Hinv".
{ iNext. iExists _, _. iFrame. iPureIntro. auto. }
wp_pures.
iApply "H$\Phi$".
iExists _. iExists _. iFrame. iModIntro. iExists $\gamma_1$. iExists $\gamma_2$.
iSplit. first eauto. iSplitL "Ha H$\gamma_1$".
- iExists _. unfold account_inv. iSplitR. eauto with iFrame. 
   iExists $\gamma_1$. iExists _. iFrame. 
-  iSplitL "Hb H$\gamma_2$". iExists _. unfold account_inv. iSplitR. eauto with iFrame. iExists $\gamma_2$. iExists _. iFrame. done. 
Qed.
\end{lstlisting}
\end{figure}

\newpage
\section{Related Work}
After working through this guide, we assume readers will want further Iris and Coq resources, a few of which we will mention here. 
\begin{itemize}
    \item To further your knowledge on Iris tactics, we refer readers to the \MYhref{https://gitlab.mpi-sws.org/iris/iris/blob/master/docs/proof_mode.md}{Iris documentation}.
    \item To further your knowledge on setting up Iris proofs in Coq, we refer readers to the \MYhref{https://gitlab.mpi-sws.org/iris/iris/-/blob/master/docs/proof_guide.md}{Iris Proof Guide}.
    \item To further your knowledge on HeapLang tactics, we refer readers to the \MYhref{https://gitlab.mpi-sws.org/iris/iris/-/blob/master/docs/heap_lang.md}{HeapLang documentation}. 
    \item For more help with installing Iris on your device, refer to this \MYhref{https://gitlab.mpi-sws.org/iris/iris}{in-depth guide}.
    \item For help inputting and outputting unicode characters throughout Iris, refer to this \MYhref{https://gitlab.mpi-sws.org/iris/iris/-/blob/master/docs/editor.md}{guide}. 
    \item For more help with installing opam on your device (before you install Iris), refer to the \MYhref{https://opam.ocaml.org/doc/Install.html}{opam documentation}.
    \item To further your knowledge on Coq and Coq tactics, we refer readers to the \MYhref{https://coq.inria.fr/refman/index.html}{Coq documentation}.
    \item For more help with installing Coq on your device, refer to the \MYhref{https://coq.inria.fr/opam-using.html}{Coq Proof Assistant documentation}. 
    \item For more details on separation logic in relation to Iris and concurrent data structures, we refer readers to the following papers: \MYhref{https://www.cambridge.org/core/services/aop-cambridge-core/content/view/26301B518CE2C52796BFA12B8BAB5B5F/S0956796818000151a.pdf/iris_from_the_ground_up_a_modular_foundation_for_higherorder_concurrent_separation_logic.pdf}{Iris from the ground up}, \MYhref{https://iris-project.org/tutorial-pdfs/iris-lecture-notes.pdf}{Lecture Notes on Iris: Higher-Order Concurrent Separation Logic}, and \MYhref{https://iris-project.org/pdfs/2017-esop-iris3-final.pdf}{The Essence of Higher-Order Concurrent Separation Logic}. 
    \item For more proof examples, many of which are advanced, refer to the \MYhref{https://gitlab.mpi-sws.org/iris/examples/-/tree/master/theories/lecture_notes}{Iris project note files}. 
    \item For other Iris and Coq guides that build off of this one, we suggest the following: \MYhref{https://plv.csail.mit.edu/blog/iris-intro.html}{A brief introduction to Iris}, the \MYhref{https://iris-project.org/tutorial-material.html}{Iris Project Lecture Notes}, and once again the \MYhref{https://iris-project.org/tutorial-pdfs/iris-lecture-notes.pdf}{Lecture Notes on Iris: Higher-Order Concurrent Separation Logic}
\end{itemize}

\newpage

\bibliography{bib}
\bibliographystyle{acl_natbib}
\nocite{*}

\newpage
\iflanguage{english}%
  {\section*{Appendix}
  \addcontentsline{toc}{section}{Appendix}
  We include a brief reference for inputting unicode symbols in Vscde with the 'Generic Input Method' extension installed and configured. More details on how to complete these steps can be found in the \MYhref{https://gitlab.mpi-sws.org/iris/iris/-/blob/master/docs/editor.md}{general unicode guide for Iris}. 

\begin{figure}[H]
    \centering
  \begin{tabular}{ |p{12mm}||p{34mm}|p{20mm}|p{75mm}|  }
 \hline
 \multicolumn{4}{|c|}{Unicode Symbols for Iris} \\
 \hline
 Symbol & Name & Code Point & Vscode Generic Input Method (prefix with $\backslash$) \\
 \hline
 $\neg$   & \scriptsize{NOT SIGN}  & 0xac &   not\\
$\uparrow$ &  \scriptsize{UPWARDS ARROW} & 0x2191 & uparrow\\
 $\rightarrow$ & \scriptsize{RIGHTWARDS ARROW} & 0x2192&  to\\
 $\leftrightarrow$ & \scriptsize{LEFT RIGHT ARROW} & 0x2194 &  iff\\
 $\mapsto$ & \scriptsize{RIGHTWARDS ARROW FROM BAR} & 0x21a6& mapsto\\
 $\forall$ & \scriptsize{FOR ALL}  & 0x2200   &all\\
 $\exists$ & \scriptsize{THERE EXISTS}  & 0x2203 &ex\\
 $\emptyset$ & \scriptsize{EMPTY SET}  & 0x2205 &empty\\
 $\in$ & \scriptsize{ELEMENT OF}  & 0x2208 &in\\
 $\not \in$ & \scriptsize{NOT AN ELEMENT OF}  & 0x2209 &notin\\
 $\backslash$ & \scriptsize{SET MINUS}  & 0x2216 &-\\
 $*$ & \scriptsize{ASTERISK OPERATOR}  & 0x2217 &star\\
 $\circ$ & \scriptsize{RING OPERATOR}  & 0x2218 &comp\\
 $\land$ & \scriptsize{LOGICAL AND}  & 0x2227 &and\\
 $\lor$ & \scriptsize{LOGICAL OR}  & 0x2228 &or\\
 $\cap$ & \scriptsize{INTERSECTION}  & 0x2229 &cap\\
 $\cup$ & \scriptsize{UNION}  & 0x222a &cup\\
 $\not =$ & \scriptsize{NOT EQUAL TO}  & 0x2260 &neq\\
 $\equiv$ & \scriptsize{IDENTICAL TO}  & 0x2261 &==\\
 $\leq$ & \scriptsize{LESS THAN OR EQUAL TO}  & 0x2264 &leq\\
 $\gg$ & \scriptsize{MUCH GREATER THAN}  & 0x226b &\\
 $\preceq$ & \scriptsize{PRECEDES OR EQUAL TO}  & 0x227c &incl\\
 $\subseteq$ & \scriptsize{SUBSET OF OR EQUAL TO}  & 0x2286 &subseteq\\
 \hline 
 
\end{tabular}
\end{figure}  

\begin{figure}[H]
    \centering
  \begin{tabular}{ |p{12mm}||p{34mm}|p{20mm}|p{75mm}|  }
 \hline
 \multicolumn{4}{|c|}{Unicode Symbols for Iris} \\
 \hline
 Symbol & Name & Code Point & Vscode Generic Input Method (prefix with $\backslash$) \\
 \hline
 $\vdash$ & \scriptsize{RIGHT TACK}  & 0x22a2 &ent\\
 $\top$ & \scriptsize{DOWN TACK}  & 0x22a4 &top\\
 $\cdot$ & \scriptsize{DOT OPERATOR}  & 0x22c5 &mult\\
 $\ulcorner$ & \scriptsize{TOP LEFT CORNER}  & 0x231c &lc\\
 $\urcorner$ & \scriptsize{TOP RIGHT CORNER}  & 0x231d &rc\\
 $\boxed{}$ & \scriptsize{WHITE SQUARE}  & 0x25a1 &box\\
 $\triangleright$ & \scriptsize{WHITE RIGHT POINTING TRIANGLE}  & 0x25b7 &later\\
 $\diamond$ & \scriptsize{WHITE DIAMOND}  & 0x25c7 &diamond\\
 $\bullet$ & \scriptsize{BLACK CIRCLE}  & 0x25cf &auth\\
 $\bigcirc$ & \scriptsize{LARGE CIRCLE}  & 0x25ef &frag\\
 $\checkmark$ & \scriptsize{CHECK MARK}  & 0x2713 &valid\\
 $\xMapsto{}{}$ & \scriptsize{RIGHTWARDS DOUBLE ARROW}  & 0x2907 &\\
 \textonehalf & \scriptsize{VULGAR FRACTION ONE HALF}  & 0xbd &\\
 $_m$ & \scriptsize{LATIN SUBSCRIPT SMALL LETTER m}  & 0x2098 &\_m\\
 $ö$ & \scriptsize{LATIN LETTER O WITH DIAERESIS}  & 0xf6 &"o\\
 $\gamma$ & \scriptsize{GREEK SMALL LETTER GAMMA}  & 0x3b3 &gamma\\
 $\epsilon$ & \scriptsize{GREEK SMALL LETTER EPSILON}  & 0x3b5 &epsilon\\
 $\lambda$ & \scriptsize{GREEK SMALL LETTER LAMBDA}  & 0x3bb &lambda\\
 $\Sigma$ & \scriptsize{GREEK CAPITAL LETTER SIGMA}  & 0x3a3 &Sigma\\
 $\varphi$ & \scriptsize{GREEK SMALL LETTER PHI}  & 0x3c6 &phi\\
 $\Phi$ & \scriptsize{GREEK PHI SYMBOL}  & 0x3d5 &\\
 $\phi$ & \scriptsize{GREEK CAPITAL LETTER PHI}  & 0x3a6 &Phi\\
 $\psi$ & \scriptsize{GREEK CAPITAL LETTER PSI}  & 0x3a8 &Psi\\
 \hline 
\end{tabular}
\end{figure}  
}

\end{document}